\documentclass[conference]{IEEEtran}

\usepackage{cite}
\usepackage{amsmath,amssymb,amsfonts}
\usepackage{algorithmic}
\usepackage{graphicx}
\usepackage{textcomp}
\usepackage{xcolor}
\usepackage{url}
\usepackage{csquotes} %

\usepackage{siunitx}
\def\BibTeX{{\rm B\kern-.05em{\sc i\kern-.025em b}\kern-.08em
    T\kern-.1667em\lower.7ex\hbox{E}\kern-.125emX}}

\usepackage{subfig}

\usepackage[nolist]{acronym}

\usepackage[english]{babel}

\newcommand{\mysubsubsubsection}[1]{
	\smallskip
	\noindent\textit{\textbf{#1\enskip}}
}

\newcounter{steps}
\newcommand*{\thestep}{%
	\arabic{steps}%
}

\newlength{\halflinewidth}
\usepackage{ifthen}

\newboolean{cameraready}
\setboolean{cameraready}{true}

\newboolean{arxiv}
\setboolean{arxiv}{true}

\ifthenelse{\boolean{arxiv}}{
	\usepackage[frozencache=true]{minted} %
}{
	\usepackage[finalizecache=true]{minted} %
}
\setminted[c]{frame=lines,linenos,breaklines}
\setminted[gas]{frame=lines,linenos,breaklines}

\ifthenelse{\boolean{cameraready}}{
	\ifthenelse{\boolean{arxiv}}{
	}{
		\usepackage[hidelinks]{hyperref}
	}
	
}{
	\usepackage[hidelinks]{hyperref}
}

\ifthenelse{\boolean{arxiv}}{
	\usepackage[printwatermark]{xwatermark}
	\newwatermark*[firstpage,color=black!80,angle=0,xpos=-50,ypos=-128, align=center,fontsize=6.8,fontseries=m]{%
		\xwmminipage[width=.3\linewidth,textalign=justified]{\textcopyright~2021 IEEE. Personal use of this material is permitted. Permission from IEEE must be obtained for all other uses, in any current or future media, including reprinting/republishing this material for advertising or promotional purposes, creating new collective works, for resale or redistribution to servers or lists, or reuse of any copyrighted component of this work in other works.}}
	\newwatermark*[firstpage,color=black!80,angle=0,xpos=49,ypos=-123.5, align=center, fontsize=6.8,fontseries=m]{%
		\xwmminipage[width=.3\linewidth,textalign=justified]{\textit{Note:} This is the authors' version of the article accepted for publication at the 2021 Workshop on Fault Detection and Tolerance in Cryptography (FDTC).}}
	\newwatermark*[pages=2-17,color=black!80,angle=0,xpos=50,ypos=-120,align=left,fontsize=6.8,fontseries=m]{\textcopyright~2021 IEEE}
}{}

\begin{document}

\title{
The Forgotten Threat of Voltage Glitching:\\ A Case Study on Nvidia Tegra X2 SoCs
}

\author{
\ifthenelse{\boolean{cameraready}}{	
	\IEEEauthorblockN{Otto Bittner\IEEEauthorrefmark{1}\textsuperscript{\textsection},
		Thilo Krachenfels\IEEEauthorrefmark{1}\textsuperscript{\textsection},
		Andreas Galauner\IEEEauthorrefmark{2} and
		Jean-Pierre Seifert\IEEEauthorrefmark{1}\IEEEauthorrefmark{3}}
	\IEEEauthorblockA{\IEEEauthorrefmark{1} Technische Universit\"at Berlin, Chair of Security in Telecommunications\\
	}
	\IEEEauthorblockA{\IEEEauthorrefmark{2} Independent Researcher\\
	}
	\IEEEauthorblockA{\IEEEauthorrefmark{3} Fraunhofer SIT\\
	}
}{
	\IEEEauthorblockN{Anonymous Author(s)}
}
}

\maketitle

\ifthenelse{\boolean{cameraready}}{
\begingroup\renewcommand\thefootnote{\textsection}
\footnotetext{These authors contributed equally to this work.}
\endgroup
}{}

\ifthenelse{\boolean{cameraready}}{
	\ifthenelse{\boolean{arxiv}}{
		\thispagestyle{plain}
		\pagestyle{plain}
	}{}
}{
	\thispagestyle{plain}
	\pagestyle{plain}
}

\begin{abstract}

Voltage \ac{FI} is a well-known attack technique that can be used to force faulty behavior in processors during their operation.
\emph{Glitching} the supply voltage can cause data value corruption, skip security checks, or enable protected code paths.
At the same time, modern \acp{SOC} are used in security-critical applications, such as self-driving cars and autonomous machines.
Since these embedded devices are often physically accessible by attackers, vendors must consider device tampering in their threat models.
However, while the threat of voltage \ac{FI} is known since the early 2000s, it seems as if vendors still \emph{forget} to integrate countermeasures.
This work shows how the entire boot security of an \emph{Nvidia \ac{SOC}}, used in Tesla's autopilot and Mercedes-Benz's infotainment system, can be circumvented using voltage \ac{FI}.
We uncover a hidden bootloader that is only available to the manufacturer for testing purposes and disabled by fuses in shipped products.
We demonstrate how to re-enable this bootloader using \ac{FI} to gain code execution with the highest privileges, enabling us to extract the bootloader's firmware and decryption keys used in later boot stages.
Using a hardware implant, an adversary might misuse the hidden bootloader to bypass trusted code execution even during the system's regular operation.

\end{abstract}


\begin{acronym}
    \acro{VR}[VR]{voltage regulator}
    \acro{SOC}[SoC]{system on a chip}
    \acroplural{SOC}[SoCs]{systems on a chip}
    \acro{PCB}[PCB]{printed circuit board}
    \acro{IC}[IC]{integrated circuit}
    \acro{PC}[PC]{program counter}
    \acro{EM}[EM]{electromagnetic}
    \acro{SPI}[SPI]{Serial Peripheral Interface}
    \acro{ROM}[ROM]{read-only memory}
    \acro{iROM}[iROM]{internal read-only memory}
    \acro{OEM}[OEM]{original equipment manufacturer}
    \acro{BR}[BR]{BootRom}
    \acro{MB1}[MB1]{MicroBoot1/nvboot}
    \acro{MB2}[MB2]{TegraBoot/TBoot-BPMP}
    \acro{FI}[FI]{fault injection}
    \acro{SCA}[SCA]{side-channel analysis}
    \acro{UART}[UART]{Universal Asynchronous Receiver Transmitter}
    \acro{RCM}[RCM]{Recovery Mode}
    \acro{DUT}[DUT]{device under test}
    \acro{TRM}[TRM]{Technical Reference Manual}
    \acro{BPMP}[BPMP]{Boot and Power Management Processor}
    \acro{AON}[AON]{Always On}
    \acro{PMIC}[PMIC]{power management integrated circuit}
    \acro{BCT}[BCT]{Boot Configuration Table}
    \acro{CCPLEX}[CCPLEX]{CPU complex}
    \acro{FSKP}[FSKP]{Factory Secure Key Provisioning}
    \acro{THR}[THR]{Transmit Holding Register}
    \acro{LSR}[LSR]{Line Status Register}
    \acro{FI}[FI]{fault injection}
    \acro{ROT}[RoT]{root of trust}
    \acro{IP}[IP]{intellectual property}
    \acro{AES}[AES]{Advanced Encryption Standard}
    \acro{CBC}[CBC]{Cipher Block Chaining}
    \acro{ECB}[ECB]{Electronic Code Book}
    \acro{IV}[IV]{initialization vector}
    \acro{FEK}[FEK]{Fuse Encryption Key}
    \acro{FREK}[FREK]{Fuse ROM Encryption Key}
    \acro{L4T}[L4T]{Linux for Tegra}
    \acro{PLL}[PLL]{phase-locked loop}
    \acro{GPIO}[GPIO]{general-purpose I/O}
    \acro{QSPI}[QSPI]{Queued Serial Peripheral Interface}
    \acro{eMMC}[eMMC]{embedded MultiMediaCard}
    \acro{FPGA}[FPGA]{field-programmable gate array}
    \acro{MOSFET}[MOSFET]{metal–oxide–semiconductor field-effect transistor}
    \acro{ECU}[ECU]{electronic control unit}
    \acro{PWM}[PWM]{pulse-width modulation}
    \acro{ODM}[ODM]{original design manufacturer}
\end{acronym}

\acresetall
\section{Introduction}
Modern \acp{SOC} are used in various applications where chip security directly relates to human safety, such as self-driving cars and autonomous machines. %
To provide secure and authenticated operation of the system, \acp{SOC} contain a trust anchor in the form of a secured and tamper-proof bootloader.
That bootloader's task is to allow the execution of only authenticated firmware components, protect \ac{IP}, set security configurations, and transfer control to less privileged boot stages.
Typically, dedicated processor cores on the \ac{SOC} act as the \ac{ROT} and are responsible for these delicate tasks.

In order to function as intended, \acp{IC} need to be operated under specified conditions, for instance, within the rated supply voltage, clock stability, temperature, and electromagnetic field ranges~\cite{bar-el_sorcerer_2006}.
This dependency can be misused to force faulty behavior during the chip's operation.
Hence, the susceptibility of electronic circuits to coincidentally or deliberately injected faults has been studied for some decades.
Especially in the smartcard field, \ac{FI} attacks used to extract secrets from cryptographic algorithms %
were investigated and counteracted around the year 2000~\cite{koemmerlin_design_1999, boneh_importance_2001, bar-el_sorcerer_2006}.
Next to the intended corruption of data values, faults can be used to skip security checks, enter protected code paths, or gain code execution~\cite{timmers_controlling_2016, lu_injecting_2019}. 
During the past years, attacks against microcontrollers and \acp{SOC} using laser-based~\cite{vasselle_laserinduced_2020} and electromagnetic~\cite{abdellatif_silicontoaster_2020, trouchkine_electromagnetic_2021} \ac{FI} have been presented.
While these techniques offer high accuracy in targeting a specific part of the chip, they also require comparatively sophisticated setups.

A simpler approach to inject faults into the system is voltage \ac{FI}, where the supply voltage is over- or undervolted for a short moment~\cite{djellid-ou_supply_2006}.
The technique is also referred to as \emph{voltage glitching}.
Even tough the technique is simpler to execute, it is repeatedly used to attack modern targets, like the Nvidia Tegra X1 \ac{SOC}~\cite{galauner_glitching_2018}.
By injecting faults into security registers, the bootloader's code was extracted and used to find a firmware bug enabling unauthenticated code execution on the system.
The fact that such simple attacks are still possible, 20 years after smartcards have been hardened against \ac{FI}, suggests that chip manufacturers seem to have ignored this threat.
Although protecting \acp{SOC} can be more difficult due to multiple power domains, complex power trees, and higher power consumption, manufacturers should implement protections against obvious and known attacks that can break the security of the system or even the entire device family.

The subsequent generation of the above mentioned \ac{SOC}, the Nvidia Tegra X2 (codename "Parker"), is used in safety-critical applications, such as for the Nvidia autonomous driving units DRIVE PX Parker AutoChauffeur and AutoCruise~\cite{nvidiacor_autonomous_2021} used in Tesla cars~\cite{nvidiacor_tesla_2016}, 
or the infotainment system in Mercedes~\cite{daimlerag_glance_2018} and Hyundai~\cite{abuelsamid_hyundai_2020} cars.
In our work, we address the following question:
\emph{Is the Tegra X2 \ac{SOC} susceptible to voltage \ac{FI} as well, or was it a bad coincidence that the previous generation was vulnerable?}

\noindent\textbf{Our contribution}\enskip\enskip
In this work, we indeed show that the processor acting as \ac{ROT} on the Tegra X2 \ac{SOC} is susceptible to voltage \ac{FI}.
We demonstrate how an attacker can gain code execution in the secure zone of the boot processor with only cheap and readily available equipment.
This capability allows us to extract the content of the \ac{iROM}, containing the first bootloader and key material used for decrypting the code of later boot stages.
This endangers the \ac{IP} of \acp{OEM} and can defeat trusted code execution.
To this end, we explain how a hardware implant can permanently manipulate the \ac{ROT}.
Since our attack cannot be easily prevented by firmware patches, we propose and discuss potential mitigations against voltage \ac{FI} attacks for future chip generations.

\smallskip
\noindent\textbf{Responsible Disclosure}\enskip
We responsibly disclosed our findings to Nvidia, including our experimental setup and parameters.
Nvidia reconstructed our experiments and confirmed that fault injection impacts the tested Tegra Parker \ac{SOC} and earlier chips.
According to them, all newer Tegra \acp{SOC} would contain countermeasures to mitigate these types of attacks.
Furthermore, they proposed countermeasures to reduce the effectiveness of voltage fault injection on vulnerable chips, which we discuss in Section~\ref{sec:discussion:mitigations:existing}.
\section{Background -- Voltage Fault Injection\label{sec:background_FI}}

\Acp{IC} need to be operated under the specified conditions to function as intended, e.g., within rated supply voltage, clock stability, temperature, and electromagnetic field ranges~\cite{bar-el_sorcerer_2006}.
This dependency can be misused to force faulty behavior during the chip's operation.
Short supply voltage variations, introduced by glitches on the supply voltage line, can produce computational errors in CMOS circuits~\cite{djellid-ou_supply_2006}.
Examples of such errors are memory bit flips, corrupted instructions, and jumping over instructions in a microprocessor.
If these errors are forced during the execution of cryptographic algorithms, information about the secret key or plaintext might be leaked~\cite{koemmerlin_design_1999, boneh_importance_2001, bar-el_sorcerer_2006}.
On the other hand, faults can be used to skip security checks, enter protected code paths, or gain code execution~\cite{timmers_controlling_2016, lu_injecting_2019}.
Voltage \ac{FI} is a well-studied field, especially due to the low cost of setups.
Open-source frameworks, such as the ChipWhisperer~\cite{oflynn_chipwhisperer_2014}, provide easy access to both hardware and software to conduct attacks.
A recent study shows that the shape of the voltage glitch can improve the attack performance, i.e., reduce the time until a successful glitch is observed~\cite{bozzato_shaping_2019}.

Depending on the design of the target, different approaches can be used to inject faults into the supply voltage rail.
If the voltage is supplied externally to the \ac{PCB}, an external power supply can introduce glitches through that interface.
If the voltage is generated directly on the \ac{PCB} using a voltage regulator, the injection of glitches becomes more complex.
On the other hand, on more advanced systems, such as \acp{SOC}, the voltage regulators typically offer communication interfaces to adjust the voltage on demand.
These interfaces, if not adequately protected, can be leveraged to inject voltage glitches~\cite{chen_voltpillager_2020}.
In some cases, the interface is even accessible via software~\cite{murdock_plundervolt_2020, qiu_voltjockey_2020}.

\begin{figure}[tb]
    \centering
    \includegraphics[width=.95\linewidth]{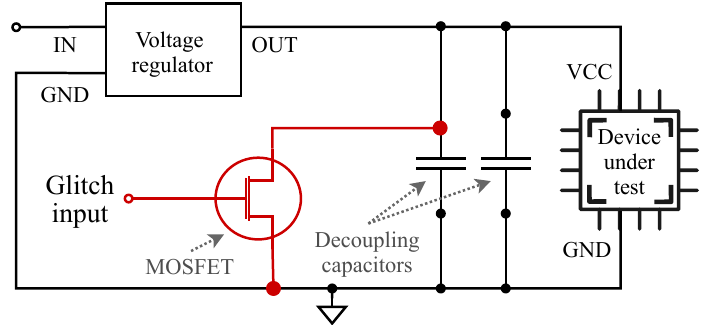}
    \caption{Schematic of a crowbar circuit (red) used to inject supply voltage glitches.}
    \label{fig:background:crowbar}
\end{figure}

Another alternative is to inject glitches using a so-called crowbar circuit.
The idea is to create a short circuit between the voltage line and GND, effectively enforcing a voltage drop~\cite{oflynn_fault_2016}.
Fig.~\ref{fig:background:crowbar} shows a schematic of such a setup.
A transistor acting as a switch -- typically an n-channel \ac{MOSFET} -- is connected between the supply voltage input (VCC) of the \ac{DUT} and GND.
To reduce noise on the supply voltage rail, \ac{PCB} designers place so-called decoupling capacitors close to the \ac{DUT}.
Their connection to VCC offers a good point for soldering the \ac{MOSFET}.
As the decoupling capacitors might reduce the effectiveness of the voltage drop, desoldering them can be beneficial to achieve shorter glitches.
During the glitch, a high short circuit current will be flowing through the \ac{MOSFET}, effectively pulling the VCC voltage close to the GND level.
It should be noted that modern \acp{SOC} typically have more than one power domain, which can complicate finding the correct rail to inject the glitch.

\section{Related Work}

As mentioned in the introduction, there is a multitude of examples for successful attacks on computational systems using fault injection.
In the following, we will introduce some of those examples that relate to our work more closely.

\subsection{Impactful Fault Injection}
With the continuous digitalization of cars, a huge market that requires secure embedded devices is the automotive sector.
In 2018, researchers from Riscure and VU University Amsterdam showed how they successfully extracted firmware from secured car \acp{ECU} using voltage glitching~\cite{milburn_ecu_glitches_2018}.
Furthermore, the study shows how the recovered firmware blobs can be emulated to quickly gain insights into the inner working of the firmware and potentially find vulnerabilities that can be exploited from software.
Similarly, the authors of~\cite{herrewegen_fill_2021} investigated vulnerabilities in microcontroller bootloaders that can be exploited by voltage glitching.
Using dynamic and static analysis of the targeted binary, they show how even a multi-glitch attack can be applied on a common microcontroller's bootloader.

In the past, also gaming consoles have been subject to repeated scrutiny from hacker groups.
There exist exploits to gain code execution on both the PlayStation 3~\cite{ps3_hack} and the Xbox 360~\cite{xbox_hack}.
Remarkably enough, for both consoles, the respective hack leverages fault injection in some way.
For the PS3, a write to a memory bus was manipulated to skip the de-allocation of a particular memory region.
This could subsequently be leveraged to gain read/write access to the hypervisor's page table, which gives the attacker full control over the system.
In the case of the Xbox 360, the attackers realized that the device's processor does not fully reset if the reset pin is only asserted for a very short amount of time.
This insight was used to skip the signature check for the second bootloader stage by pulsing the reset pin for \SI{100}{\ns}.
Both of these examples showcase how otherwise very tight security measures can be bypassed entirely using fault injection.

\subsection{Previous Tegra Generations}
When the gaming company Nintendo released their newest handheld gaming console called ``Nintendo Switch'' in 2017, initial teardowns and analyses of the hardware suggested that the \ac{SOC} employed on the platform could be an Nvidia Tegra X1 \cite{switch_tx1_dieshots}. 
As with previous gaming consoles, the Switch was investigated by several hacker groups.
Their research aimed to gain the capability to execute code on the platform, and therefore, to bypass the mandatory code signing required by Nintendo. 
While the researchers exploited typical software targets like the embedded Webkit-based browser and interfaces to the operating system kernel at first, the \ac{BR}, which enforces the root of trust, later became a target as well~\cite{szekely_qyriad_2021}. 

Compromising this part of the boot chain gives an attacker full control over the following boot process, rendering any cryptographic security measures ineffective. 
Additionally, the \ac{BR} is not patchable after production, making it impossible for Nintendo or Nvidia to fix discovered flaws. 
Since the \ac{BR} is \ac{IP} of Nvidia, it is not publicly available.
Therefore, the \ac{BR} first needed to be leaked from the \ac{SOC} for further analysis.
By default, due to the read protection, this is not possible.
Just before execution is passed from the \ac{BR} to the first cryptographically verified bootloader, the \ac{BR} activates a read protection for itself. 
Multiple independent security researchers successfully applied voltage glitching to skip the instruction that sets the read-lock bit in a control register \cite{galauner_glitching_2018, rousseltarbouriech2019methodically}.
Others used the same technique to skip the signature verification of executed bootloaders entirely \cite{console_security_switch}.
\section{Attack Approach\label{sec:attack_approach}}

\subsection{Threat Model}
The attacks described in the previous section already show that voltage \ac{FI} can break consumer products and enable unintended usage.
However, these kinds of attacks would even pose a physical threat if conducted on a platform used in security-critical applications.
The attacks have in common that the capability to execute unauthenticated code is gained.
In the case of an autonomous driving unit, an adversary could alter the firmware to tamper with, e.g., how the car reacts to human obstacles.
Even if only the cockpit display is tampered with, wrong speed values could be displayed, potentially endangering passengers and pedestrians.

Building on this threat model, we assume an attacker with physical access who wants to gain capabilities to execute unauthenticated code by applying voltage glitching.
The attacker might have several options to achieve this goal.
Firstly, they could extract the device's firmware to search it for software bugs that allow code execution.
Secondly, the attacker could fault routines that check the authenticity of code before execution.
Finally, they could directly gain code execution with elevated privileges by re-enabling a debug or testing interface.
Although the last two approaches do not allow permanent unauthenticated code execution, the attacker could use a hardware implant to provoke the fault whenever needed.
In the following, we describe how an attacker might proceed when trying to conduct either of these three approaches.

\subsection{Attack Procedure\label{sec:attack_approach:procedure}}
We identified five steps that likely have to be taken when applying any sort of \ac{FI} attacks.

\mysubsubsubsection{Step \refstepcounter{steps}\thestep\label{step_1}: Determining the feasibility of \ac{FI}}
Ideally, the attacker has some means of executing code on the target CPU. 
This can either be achieved by a software exploit that lets the attacker execute arbitrary code in an otherwise locked-down environment or by getting access to development hardware that allows code execution for development purposes. %
The executed code can then be used to test if glitches applied to multiple different voltage rails, clock inputs, reset lines, or similar external control lines might affect the correctness of instructions executed by the CPU. 
The easiest way to test for this is to build tight endless-loops that constantly add numbers and output them in short succession. 
If the code behaves correctly, the arithmetic results printed by the CPU need to be correct all the time. 
If an applied glitch changes the result of an arithmetic operation, it is a strong indicator that the CPU has experienced a fault caused by the applied glitch.

For being able to inject faults into the \ac{DUT}, some preparation needs to be done.
Depending on the target device, that involves finding the right voltage rail, clock input, or reset pin to apply a glitch.
If a voltage rail should be glitched, removing capacitors on that rail to achieve a sharper glitch pulse and attaching devices like a \ac{MOSFET} for shorting supply rails to GND is necessary, cf. Section~\ref{sec:background_FI}.
Using the prepared hardware and software setup, fault injection can now be performed to evaluate its effect on the test code.

\mysubsubsubsection{Step \refstepcounter{steps}\thestep\label{step_2}: Identifying the \ac{FI} target and a success indicator}
After determining the feasibility of \ac{FI} itself, locations in the actual code that should be glitched need to be identified. 
Typically, a glitch causes an arithmetic result to be wrong, writebacks to memory or registers to fail, or instructions to be skipped entirely. 
Depending on how the code was written and how the compiler applied optimizations, routines like cryptographic signature checks can be circumvented by skipping the branch to the code path handling a failed signature verification. 
If this branch instruction can be skipped, the execution continues in the success path and code execution can be gained although the verification failed. 
For example, in Listing~\ref{listing:glitch_signature_check}%
, a glitch could potentially skip the \verb|cbz|-instruction in line~4 to make the CPU jump into the code that failed authentication. 
If \ac{FI} should prevent the activation of a memory read-protection, either the instruction setting the protection or the writeback into the control register itself can be faulted.
\begin{listing}[tb]
	\centering
	\begin{minipage}{.85\linewidth}
	\usemintedstyle{pastie}
\begin{minted}
{gas}
    push    {fp, lr}
    bl      load_further_code
    bl      sig_verify
    cbz     r0, sigcheck_failed
    bl      call_authenticated_code
sigcheck_failed:
    bl      signify_auth_error
.hang:
    b       .hang
\end{minted}

	\end{minipage}
	\caption{Pseudocode for a signature check that can potentially be skipped. Note that the call in line 5 never returns.}
	\label{listing:glitch_signature_check}
\end{listing}

For detecting if a glitch was successful in an automated glitching setup, an externally available indication is needed.
This can be, for instance, a signal available at an external pin or specific content in a log file.
In case there is no easily available success criterion, side-channel information, e.g., from the power consumption or \ac{EM} emission, can be used for this purpose.

\mysubsubsubsection{Step \refstepcounter{steps}\thestep\label{step_3}: Finding a trigger signal}
Since a glitch needs to target the execution of a particular instruction, the attacker needs some timing anchor to measure the time until the glitch needs to be triggered. 
This timing anchor can be generated, on the one hand, on external interfaces of the device by the software under attack.
For instance, some code paths might need input like USB or \ac{UART} traffic, which needs to be generated by the attacker as well. 
In these cases, generated inputs can be used as a timing anchor. 
On the other hand, signals used for the initialization of external memories, modems, or similar external \acp{IC} are suitable as an anchor as well.
Care must be taken to avoid timing anchors where the time between the anchor and the targeted instruction jitters. 
For example, a modern \ac{SOC} contains multiple \acp{PLL} that are usually configured as soon as code execution starts. 
\acp{PLL} take a varying time to lock and produce a stable frequency output before execution can switch over to them as the new clock source.
In this case, an event happening after the code causing jitter is suited better as a stable timing anchor.

When a glitching attempt fails, the target might behave erratically, crash, or hang without any kind of output. 
In these cases, or for triggering a new glitching attempt, the target needs to be reset by the glitching hardware.

\mysubsubsubsection{Step \refstepcounter{steps}\thestep\label{step_4}: Finding glitch parameters}
After preparing the hardware and acquiring a stable timing reference, the right spot for the glitch needs to be found. 
This can be achieved by sweeping over a timespan while constantly resetting the target and applying glitches at different times until the glitch can be executed more or less reliably. 
Depending on the target, the timing reference, and the code path, it may be impossible to achieve a \SI{100}{\percent} success rate.
Success rate in this context means the number of glitches that produce the desired result in relation to the number of tries.
However, a success rate of \SI{100}{\percent} is usually not necessary.
Often a low enough time-to-success is sufficient, for which the number of tries per timeframe plays a big role.
A glitch that only works in \SI{1}{\percent} of the cases but can be tried hundreds of times per second, would be favorable to one that works in \SI{10}{\percent} of the cases but can only be tried three times per second.

\mysubsubsubsection{Step \refstepcounter{steps}\thestep\label{step_5}: Generating target payload.}
When the glitching parameters are determined, the attacker can manipulate or skip an instruction of their will.
Depending on their goal, the attacker might have to build custom payloads for the target.
For instance, if the attack enables the execution of attacker-controlled unauthenticated code, this code must be created and loaded into the \ac{DUT}. 
The payload can carry out different tasks like dumping protected code or secret key material, or chain-loading further code to run more complex payloads.

\section{Experimental Setup}

\subsection{Device Under Test: Nvidia Tegra X2} \label{subsec:dut}

The Tegra X2 (model number T186, codenamed "Parker") was introduced in January 2016 to replace the Tegra X1.
It is designed for embedded devices requiring high amounts of processing power to support gaming or machine learning applications.
It houses a Denver 2 CPU, an ARM Cortex-A57 MPCore CPU, an Nvidia GP10B Pascal GPU, and multiple Cortex-R5 processors for different functionalities.
For consumers, the X2 can be acquired from Nvidia as part of the Jetson TX2 module~\cite{nvidiacor_jetson_2017}.
The TX2 module includes the X2 \ac{SOC}, as well as external volatile and non-volatile memories, power management controllers, and a Bluetooth and WIFI modem.
We used a Jetson TX2 module together with a Jetson TX2 developer kit carrier board~\cite{nvidiacor_harness_2017} for our experiments.
The carrier board exposes an array of different signals from the TX2, including the different \ac{UART} interfaces.
To operate the TX2 module, it is plugged on the carrier board, as shown in Fig.~\ref{fig:module_vanilla_mount}, using a 400-pin SAMTEC REF-186137-03 connector.

\begin{figure}[tb]
	\includegraphics[width=\linewidth]{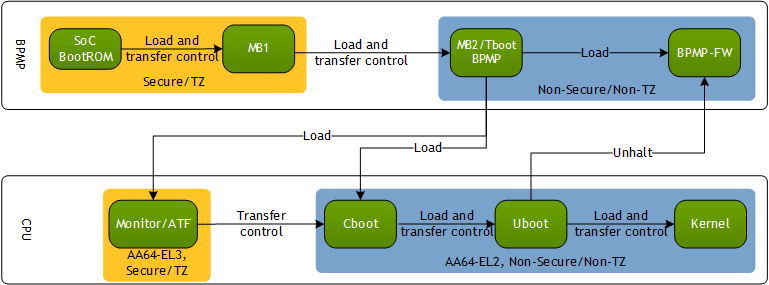}
	\caption{Tegra X2 boot flow. Figure taken from~\cite{tx2_bootflow}.}
	\label{fig:x2_bootflow}
\end{figure}

On the X2 \ac{SOC}, one of the Cortex-R5 processors, the \ac{BPMP}, is of particular interest to us as it is responsible for the initial boot process.
The \ac{BPMP} is the first processor in the boot flow and starts execution in the Secure/TZ mode (see Fig.~\ref{fig:x2_bootflow}).
In this mode, the \acf{BR} and \ac{MB1} are executed.
The \ac{BR} is hardcoded into the X2's \ac{iROM} and forms the \ac{ROT} for the \ac{SOC}.
It can only be changed slightly through a patching mechanism that uses fuse storage to apply patches to the \ac{BR}, called iPatch.
However, there is no information publicly available on this feature. 

For the X2, the privileges associated with being in Secure/TZ mode are not publicly described by Nvidia either.
While the Tegra X1 TRM gives information on how Secure/TZ mode is implemented, the X2 \ac{TRM}~\cite{x2_trm} does not.
From the processor type, we can infer that the Cortex-R5 does not implement ARM's TrustZone. 
This aligns with the fact that the \ac{BPMP} on the X1, an ARM7TDMI processor, did not implement ARM's TrustZone either.
From the previous work on the X1~\cite{galauner_glitching_2018}, we can assume that the \ac{iROM} should only be fully readable in the Secure/TZ mode.
However, we only developed a closer understanding of the mode while working with the X2.
Our findings are described in the Sections~\ref{subsubsec:unearthing} and~\ref{subsubsec:analyzing_irom}.

After startup, the \ac{BR} loads the second boot stage called \ac{MB1}, which is provided by Nvidia as an encrypted and signed binary.
It can only be changed by Nvidia, which is advertised as a feature to allow for post-production modifications of the boot flow.
\ac{MB1} resides in external eMMC memory.
Once the \ac{BR} successfully verifies and decrypts \ac{MB1}, it yields control to it.
During the execution of \ac{MB1}, more devices and cores are brought up.
For \acp{OEM}, Nvidia also offers possibilities to implement secure boot features by attesting the integrity of the next boot stage, \ac{MB2}, and possibly encrypting it.

As control is yielded to the \ac{MB2} stage, privileges are dropped to the Non-Secure/Non-TZ mode.
One task of \ac{MB2} is to hand over control to the \ac{CCPLEX} (see the lower half in Fig.~\ref{fig:x2_bootflow}).
The boot stages that are executed on the \ac{CCPLEX} are not relevant for us as they do not have access to the protected \ac{iROM} of the \ac{BPMP}.
Since \ac{MB2} is modifiable by \acp{OEM}, no security features are keeping us from changing the \ac{MB2} binary and executing our own code.
However, as only \acp{OEM} are intended to introduce modifications at this level, there is no official documentation on how to build a working binary.

\subsection{Fault Injection Setup}

\begin{figure}
    \centering
    \includegraphics[width=\linewidth]{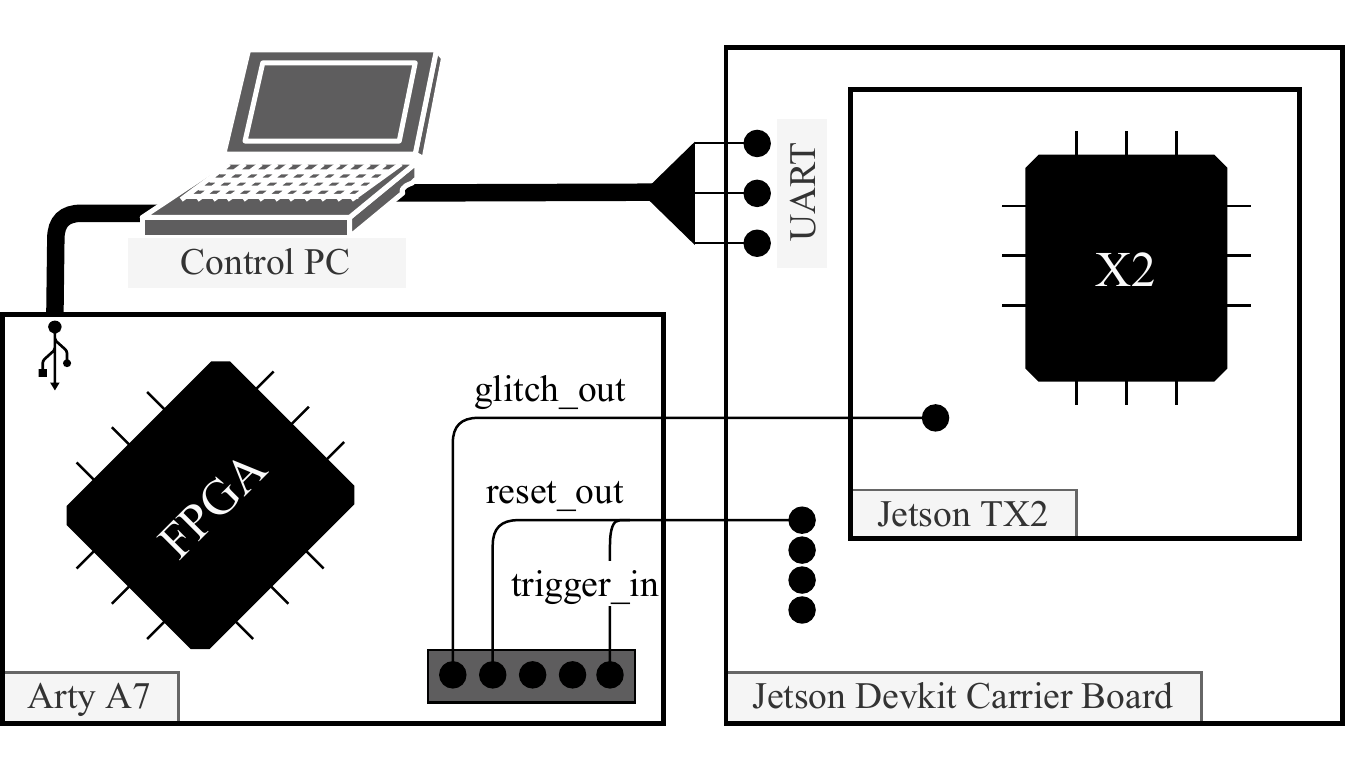}
    \captionsetup{skip=-1pt}
    \caption{Overview of the FI setup.}
    \label{fig:setup_overview}
\end{figure}

The three main components of our setup are (a) a control PC, (b) a \ac{FPGA}, and (c) the X2 \ac{SOC}, see Fig.~\ref{fig:setup_overview}.
The \ac{FPGA} is a Xilinx Artix-7 XC7A35T, located on a Digilent Arty-A7 development board that provides access to the \ac{FPGA}'s \ac{GPIO} ports and offers a USB-to-\acsu{UART} bridge.
Through the \ac{UART} bridge, we can establish a communication channel between the control PC and the \ac{FPGA}.
The \ac{FPGA} is used to have exact control over the timing of the glitches.
Since that timing is on the nanosecond scale, speed is a crucial consideration.

To configure the glitch parameters, we implemented a command-based protocol on the \ac{FPGA}.
This implementation is heavily based on an open-source implementation by chip.fail~\cite{chipfail_repo}.
The project includes Verilog code for an \ac{FPGA} and the necessary Python scripts to control the \ac{FPGA} using the command-based protocol.
In our case, the Python code implements a brute force algorithm to randomly traverse the search space of glitch parameters and configure the \ac{FPGA} for each parameter combination.
The glitch parameters are (a) the glitch length and (b) the delay between the trigger signal and the beginning of the glitch activation (offset).

The control PC can also communicate with the X2 \ac{DUT} using a USB-to-\ac{UART} adapter, which allows us to read early boot logs and output from programs we run on the X2.
For resetting the X2, we use one of the reset signals exposed on the carrier board.
By connecting an \ac{FPGA} \ac{GPIO} port to the carrier board, we can reset the X2 by toggling the pin.

\begin{figure}
	\centering
    \subfloat[Module without adapter\label{fig:module_vanilla_mount}]{
        \includegraphics[width=\halflinewidth]{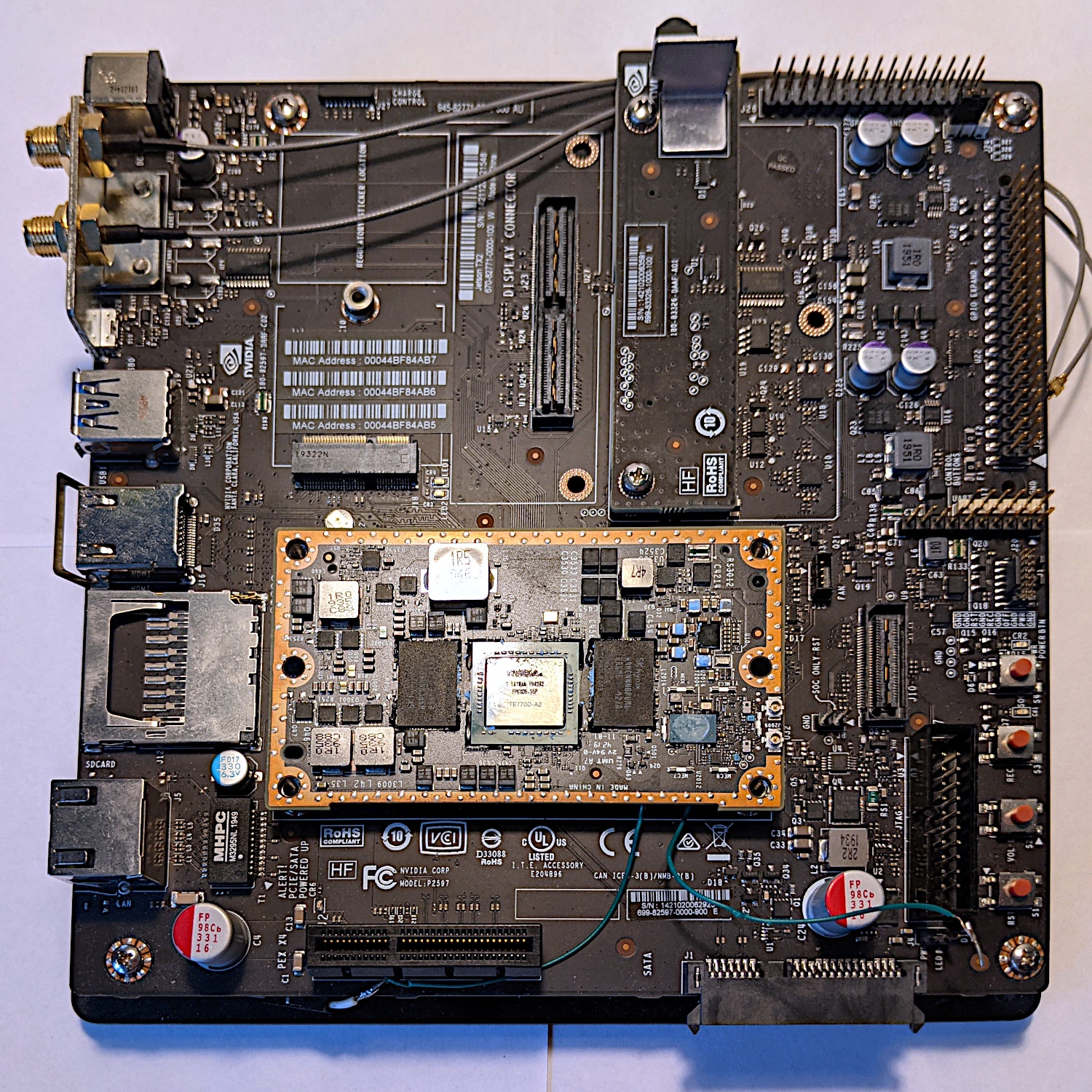}
    }\hfill
	\subfloat[Module with adapter\label{fig:module_adapter_mount}]{
		\includegraphics[width=\halflinewidth]{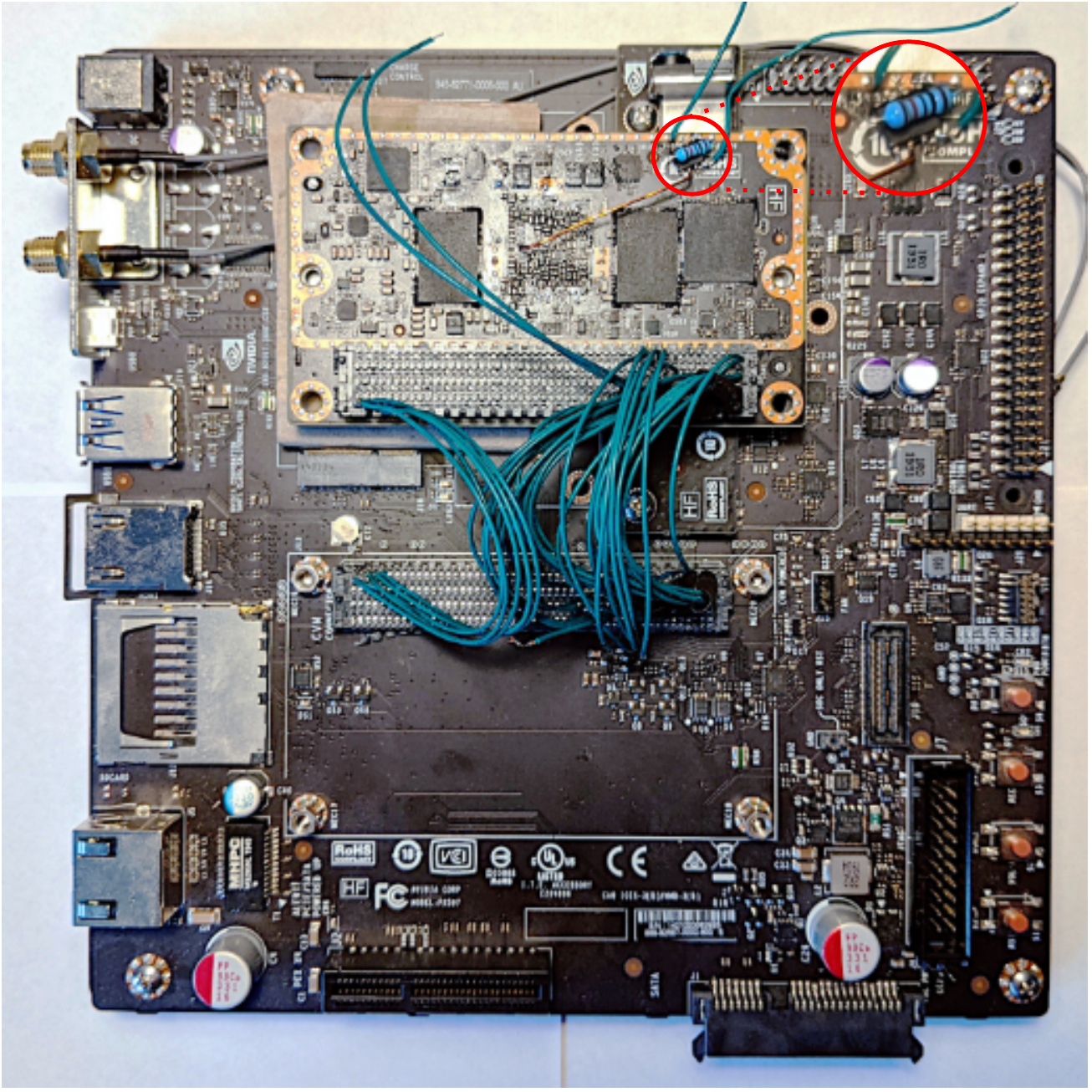}
	}
    \caption{TX2 module mounted on the developer kit carrier board.\label{fig:module}}
\end{figure}

To inject voltage glitches, we soldered an n-channel Infineon IRF8736PbF \ac{MOSFET} to the targeted voltage rail on the TX2 module with a pull-down resistor on its gate, see zoom-in of Fig.~\ref{fig:module_adapter_mount}.
The transistor's gate is controlled by a Maxim Integrated MAX4619 multiplexer that acts as a level shifter and switches between ground and \SI{5}{V}.
The multiplexer, in turn, is controlled by an \ac{FPGA} \ac{GPIO} to trigger a glitch.

\section{Results}

\subsection{Injecting Faults into Code Execution}
For testing the \ac{BPMP}'s susceptibility to \ac{FI}, we ideally need the ability to execute our own code on this processor, as described by Step~\ref{step_1} in Section~\ref{sec:attack_approach:procedure}.
In our code, we need to issue a stable trigger signal available at an external pin of the \ac{DUT}.
Since we might regularly have to reset the chip during testing, our code should be executed as early as possible in the boot flow.
Furthermore, we need to find the correct voltage rail on the \ac{PCB} that correlates with the \ac{BPMP}.

\subsubsection{MB2 Code Execution}
In Section~\ref{subsec:dut}, we mentioned that the \ac{MB2} boot stage is intended to be modified by \acp{OEM}.
However, no public documentation is available on how to do this.
Thus, we had to develop a working binary for the Cortex-R5 processor blindly.

In order to cross-compile binaries for the \ac{BPMP}, we used the GNU Arm Embedded Toolchain, version 9.3.1.
To set up a C runtime environment for \ac{MB2}- the \verb|.bss|-section must be zeroized so that uninitialized variables in the C code have a predictable default value.
Furthermore, the stack pointer must be set to the correct address before jumping into the main function.
This startup code is written in assembly and built using the ARM toolchain mentioned above.
We configured the compiler to skip all standard system startup files or libc function imports.

Since we had to define the base address for the binary explicitly, we used a custom linker script.
We found the base address by running a tool called rbasefind~\cite{gayou_sgayou_2020} on the original binary. 
It searches the binary for strings and pointers and calculates the number of pointers directing to the found strings for different base addresses.
Depending on the number of pointers that point to a string, the tool recommends the most likely base address candidates.
In our case, the tool showed the address \verb|0x52000000| as the most promising candidate.
Loading the original binary into Ghidra using this address indicated its correctness.
This was confirmed by successful code execution in \ac{MB2}, which gave us control over the \ac{BPMP} in the Non-Secure/Non-TZ mode.
Most importantly, at this point, we are able to communicate with the control PC via \ac{UART} and define trigger signals using \acp{GPIO}.

\subsubsection{Finding the BPMP's Voltage Rail}
Finding the correct voltage rail to inject glitches was the next challenge.
There is only a high-level description of the TX2 module available, mentioning information necessary for hardware designers to integrate the module on their own carrier boards. 
The full power tree is not publicly documented. 
However, several switching voltage regulators can be identified by their large inductors and capacitors when looking at the module.
Three voltage regulators are uP1666Q buck controllers that support the Nvidia OpenVReg Type 2+1 PWMVID feature~\cite{nvidiacor_openvreg_2011}.
They allow dynamic voltage control via a \ac{PWM} signal, which the X2 \ac{SOC} generates.

\begin{figure}
	\centering
	\includegraphics[width=.75\linewidth]{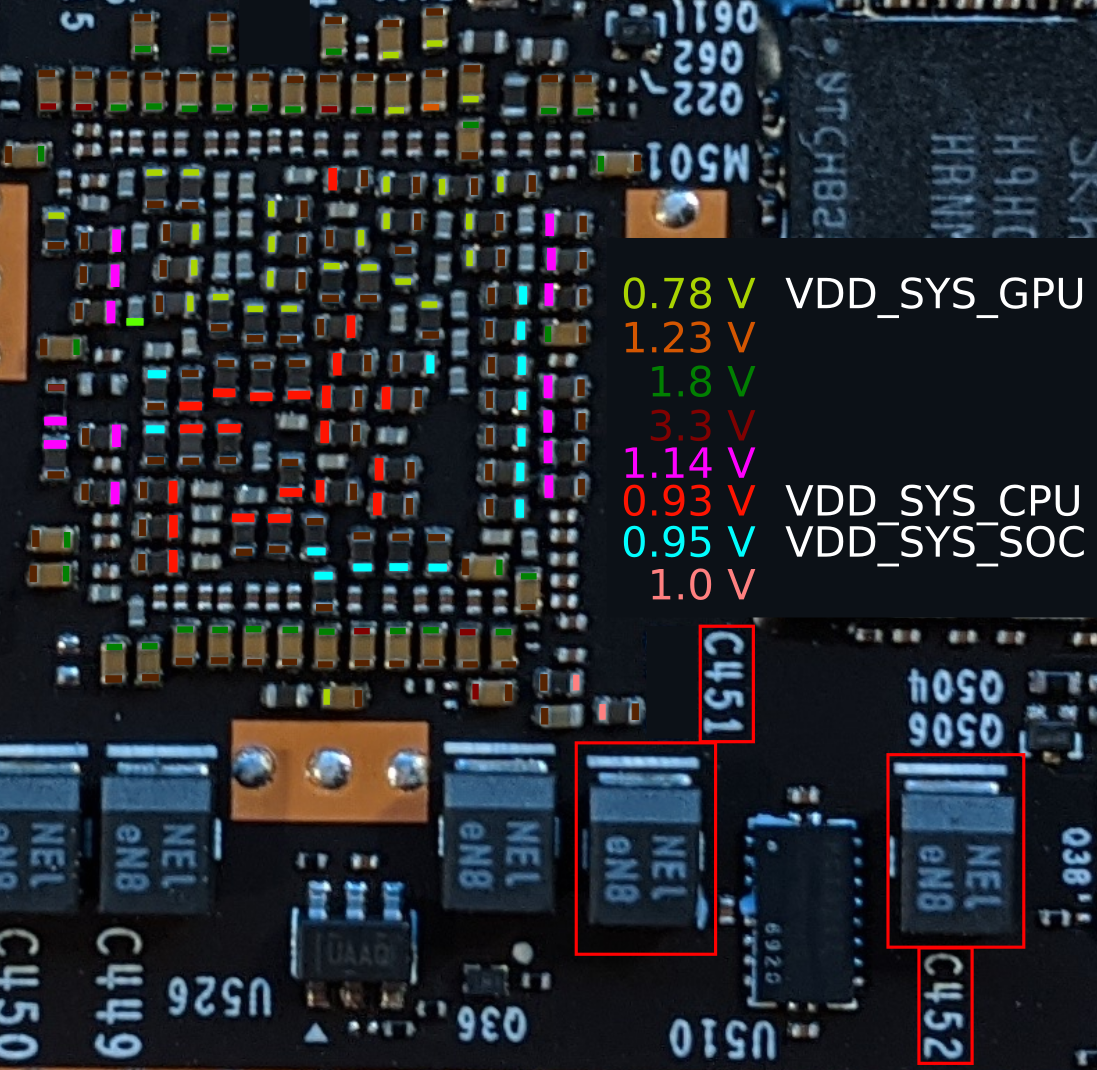}
	\caption{The different voltages on decoupling capacitors below the SoC.}
	\label{fig:caps}
\end{figure}

Since \ac{MB1} starts most of the processors in the X2, there needs to be a way of controlling the \ac{PWM} signals.
The way developers can influence the behavior of \ac{MB1} is the \ac{BCT}.
This table includes configuration values for devices that \ac{MB1} initializes and is built by a software toolchain included in Nvidia's \ac{L4T} package~\cite{nvidia_l4t}.
The Driver Package Development Guide~\cite{x2_driver_guide} describes how the voltage configuration through the \ac{BCT} is done in detail.
From the \ac{L4T} package files that control the configuration, we can learn the default voltage levels and within which boundaries the rail should be operated.
Since the three rails are configured with different voltage levels, we can map the rail names \texttt{VDD\_\allowbreak SYS\_\allowbreak SOC} (\SI{0.95}{\V}), \texttt{VDD\_\allowbreak SYS\_\allowbreak CPU} (\SI{0.78}{\V}), and \texttt{VDD\_\allowbreak SYS\_\allowbreak GPU} (\SI{0.93}{\V}) to the respective voltage regulators on the \ac{PCB}.
Now knowing which voltage level correlates with which power domain, we can probe the decoupling capacitors present on the backside of the \ac{PCB} below the \ac{SOC}, see Fig.~\ref{fig:caps}.
In order to gain access to these capacitors during operation of the chip, we built an adapter using two SAMTEC connectors to place the module in an upside-down orientation, as shown in Fig.~\ref{fig:module_adapter_mount}.

The last question remaining is which power rail supplies the \ac{BPMP}.
To this regard, another observation can be used:
The \texttt{VDD\_\allowbreak SYS\_\allowbreak SOC} rail is switched on first, while \texttt{VDD\_\allowbreak SYS\_\allowbreak CPU} and \verb|VDD_SYS_GPU| are only enabled later during the boot process. 
Since the \ac{BPMP} executes the first instructions, this leads to the conclusion that \texttt{VDD\_\allowbreak SYS\_\allowbreak SOC} is the rail supplying the \ac{BPMP}.

At this point, we desoldered all decoupling capacitors shown in Fig.~\ref{fig:caps} that are connected to \verb|VDD_SYS_SOC|.
Furthermore, we removed the electrolytic capacitors with the \ac{PCB} labels C451 and C452 (red boxes in the figure).
Subsequently, we soldered the \ac{MOSFET} used for injecting glitches to the \ac{PCB} and connected its drain pin to one of the exposed pads -- that was previously occupied by a decoupling capacitor -- using a \SI{35}{\mm} piece of enameled copper wire with a diameter of \SI{0.4}{\mm}.
This setup can also be seen in Fig.~\ref{fig:module_adapter_mount}.
Afterward, we flashed a tightly coupled endless-loop as \ac{MB2}, as described in Step~\ref{step_1} of the attack procedure.
The program continuously emits the result of an addition operation on the \ac{UART} interface after each iteration.
Using a Python script to control the \ac{FPGA}, we started glitching the X2 with increasing pulse lengths while manually triggering the glitch.
We could observe corrupted data on the \ac{UART} interface, confirming that we can enforce faulty behavior on the \ac{BPMP}.

\subsection{Understanding the X2 Secure Boot}
Following Step~\ref{step_2} of the attack procedure, we now needed to look for locations in the boot process that can potentially be exploited using \ac{FI}.
During our description of the X2 \ac{SOC}, we stated that the Secure/TZ mode is not well described in the official documentation.
In order to understand the necessary details, we relied on unofficial information that can be found online.

\subsubsection{Finding Hidden Documentation} \label{subsubsec:unearthing}
Nvidia offers header files containing memory mappings for the X2's registers as part of their \ac{L4T} package.
These mappings are also described in the X2's \ac{TRM}~\cite{x2_trm}, together with the information of which header describes which memory mapping.
Since we were interested in the \ac{BR}, we were looking for memory mappings that configure the \ac{iROM}.
The memory map lists \texttt{BPMP\_\allowbreak ATCM} as parent aperture for the \texttt{BPMP\_\allowbreak BOOTROM}, indicating that the \ac{BR} is located within \texttt{BPMP\_\allowbreak ATCM}.
When inspecting the memory mappings of the aperture, one mapping seems particularly interesting: \texttt{BPMP\_\allowbreak ATCM\_\allowbreak CFG}.
The header file describing this mapping, \texttt{arbpmp\_\allowbreak atcmcfg.h},  is missing in the \ac{L4T} package and can only be found as part of Nvidia's Sensor Processing Engine source code~\cite{tx2_l4t_spe_armiscreg_header}.
When going through the file, some register names can be recognized for their similarity to important registers on the X1, like \verb|BPMP_ATCMCFG_SB_CFG_0| and \texttt{BPMP\_\allowbreak ATCMCFG\_\allowbreak SB\_\allowbreak PIROM\_\allowbreak START\_\allowbreak 0}.

\mysubsubsubsection{Remark}
Before realizing that the header file is offered by Nvidia, we searched for it on GitHub. %
Apart from finding a repository that includes the Nvidia code, the search also uncovered a repository called "switch-bootroms"~\cite{zerospacenx_2021}.
This repository includes leaked \ac{BR} source code for the Tegra \acp{SOC} with model numbers T210 and T214, whereas T210 is the original model of the Tegra X1 (codenamed "Erista"), and T214 is an updated version, also called Tegra X1+ (codenamed "Mariko")~\cite{tegra_2021}.
The X1+ includes faster clock speeds and, judging from comments and code in the repository, is hardened against \ac{FI}.
During our investigations, access to this code massively increased our understanding of the X2. %

\subsubsection{ACCESS\_PIROM and PIROM\_START}
Looking at the header file mentioned above and going through the descriptions of the different bits in the named registers, we get an idea of how the protection mechanisms work.
Three bits/registers stand out: \verb|SECURE_BOOT|, \verb|PIROM_START|, and \verb|ACCESS_PIROM|.
Judging from the naming, default values, and the fact that similarly named bits exist on the X1, we can confidently assume their respective functionality.
The \verb|PIROM_START| register holds an address marking the beginning of the protected \ac{iROM}.
No memory locations between \verb|PIROM_START| and the end of the \ac{iROM} can be accessed unless \verb|ACCESS_PIROM| is set to 1.
Both registers can only be changed while \verb|SECURE_BOOT| is set to 1.
During \ac{MB2}, this flag is set to 0.

Furthermore, a comment found in the source code from GitHub (line 492 of \verb|nvboot_bpmp.c|) stresses the importance of preventing writes to \verb|PIROM_START| outside the Secure/TZ mode.
Changing that value to the end of the \ac{iROM} address range would allow an attacker to read the entire \ac{iROM} content.
Moreover, the comment confirms that the \verb|SECURE_BOOT| flag controls write access to \verb|PIROM_START| and \verb|ACCESS_PIROM|.

\subsubsection{Secrets in the BootROM}
For understanding the implications of leaking the \ac{BR}, we consulted the official Nvidia documentation.

\mysubsubsubsection{MB1 Decryption Key}
Looking at the X2's boot flow depicted in Fig.~\ref{fig:x2_bootflow}, we know that \ac{MB1} is decrypted by the \ac{BR} stage.
At this point, we can not know if \ac{BR} accesses fuses to retrieve the key(s) or the protected \ac{iROM} includes all necessary key material.
Therefore, it is possible that \ac{MB1} can not be decrypted by only leaking the \ac{BR}, but relevant fuse data or other protected content may also have to be leaked.

\mysubsubsubsection{Factory Secure Key Provisioning} \label{par:nvidia_fskp}
Apart from key material to decrypt the second boot stage, different sources suggest the existence of a feature called \ac{FSKP}.
While no official documentation mentions this feature for the X2, Nvidia has a patent describing it~\cite{nvidia_fskp_patent}.
The existence of the feature on the X2 is further supported by, e.g., a header file describing the respective keys~\cite{abdel_tawab_fskp_keys}.
Furthermore, we saw traces of the feature in the leaked \ac{BR} code for the previous generation X1+ \ac{SOC}.

The feature allows \ac{OEM}s to encrypt data they want to burn into fuses.
This is required when the \ac{OEM}'s threat model regards the factory -- where the fuses are burnt -- as potentially compromised.
The encrypted data can only be decrypted using the \ac{FSKP} key, that is provided by Nvidia and located in the X2's \ac{iROM}.
Since the keys are programmed into the devices during production, they can not be changed at a later point. 
There are 63 keys that Nvidia can assign, one per \ac{OEM}.
Since these keys are essentially the root of trust for an \ac{OEM}'s secure boot implementation, their secrecy is of utmost importance.

\subsection{Reverse Engineering unprotected BootRom}
\subsubsection{Dumping iROM}\label{}
Once we gained a deeper understanding of the \ac{iROM} protections using the files found on GitHub, we decided to read out the unprotected section of the \ac{iROM}.
Dumping the unprotected \ac{iROM} is done by running a loop from the address marked as \verb|BOOTROM_BASE| in the \ac{TRM}, until 128 kilobytes have been read or execution is interrupted by an exception due to unauthorized memory access.
Inside the loop, each byte is sent via \ac{UART} to the control PC.

Running this code in \ac{MB2} results in a dump of the address range \verb|0x10000 - 0x11200|.
The upper bound, \verb|0x11200|, is the value stored in \verb|PIROM_START| during \ac{MB2}.
We could successfully import the read binary into Ghidra by selecting ARMv7, little-endian, 32-bit as processor type and entering the base address \verb|0x10000| mentioned in the \ac{TRM}.
This allows us to analyze the \ac{BR} code further.

\subsubsection{Analyzing iROM} \label{subsubsec:analyzing_irom}

Execution on the Cortex-R5 starts with a \verb|Reset| exception, forcing the program counter to address \verb|0x0|.
A branch to the reset handler is located at this address, which we call \verb|reset()|.
As a first step to analyzing \verb|reset()|, we look up all memory addresses that Ghidra marks as unknown in the \ac{TRM}.
One accessed address of particular interest to us is \texttt{BPMP\_\allowbreak ATCMCFG\_\allowbreak SB\_\allowbreak PIROM\_\allowbreak START\_0}. 
We can see that the value of this address is updated from \verb|0x400| to \verb|0x2000|.
This allows the \ac{BPMP} to access the \ac{BR} for subsequent initialization.
While it may be possible to attack the store instruction with \ac{FI}, it will most likely only break the boot process.

Apart from updating this address and the initialization of multiple processor status registers, two functions are called.
We call the first function \verb|ApplyIromPatches()|, which is responsible for activating a set of patches located in the X2's fuse memory.
It enables Nvidia to update the \ac{BR} in minor ways to fix critical bugs in the boot process.
We call the second function \verb|NonSecureDispatcher()|, where an array of function pointers is used to sequentially execute multiple functions to initialize the \ac{SOC} before entering the protected address range.
Lastly, the \verb|reset()| function jumps into the protected \ac{iROM} section at address \verb|0x1200|.

Taking a closer look at the functions called by \texttt{Non\allowbreak Secure\allowbreak Dispatcher()}, the second one stands out, as it looks like the one called \texttt{NvBoot\allowbreak Main\allowbreak Nonsecure\allowbreak RomEnter()} in the X1's source code.
The function checks fuse data to determine whether or not the chip currently is in \textit{FailureAnalysisMode} or \textit{PreproductionMode}.
In case the chip is in \textit{FailureAnalysisMode}, it executes a function called \texttt{Nv\allowbreak Boot\allowbreak Uart\allowbreak Download()}.
This function is highly interesting for us, as it initializes a \ac{UART} interface, sends out a prompt, and waits for data to arrive at the interface.
Once the function has read enough data, it checks a simple checksum received together with the data.
If the checksum is correct, the data is executed as code on the \ac{BPMP} in the currently active privilege mode, i.e., the Secure/TZ mode.
The prompt on the \ac{UART} interface can serve as an indication of a successful glitch, as described in Step~\ref{step_2} of the attack procedure.

Since our X2 is booting normally, we can assume that neither of the above modes is active, and thus, the interface is not available to us.
However, this code-loading feature is only protected by the fuse checks. 
There are no further security checks as to whether the loaded code is signed by an authorized party.
Consequently, if we could manipulate the fuse check, we would likely gain code execution in the Secure/TZ mode.

\subsubsection{Fuse Check Code}
\begin{listing}[b]
	\centering
	\begin{minipage}{.85\linewidth}
		\usemintedstyle{pastie}
\begin{minted}
{gas}
    push    {fp, lr}    
    bl      is_fam
    cbz     r0, is_not_fam
is_fam_or_ppm:   
    bl      is_ppm
    cbnz    r0, exit 
    bl       NvBootUartDownload 
is_not_fam:   
    bl      is_ppm
    cmp     r0, 0  
    bne     is_fam_or_ppm
exit: 
    pop     {fp, pc}    
\end{minted}
	\end{minipage}
	\caption{Pseudocode for fuse check protecting the \texttt{Nv\allowbreak Boot\allowbreak Uart\allowbreak Download()} routine.}
	\label{listing:condition_check}
\end{listing}

In order to understand whether or not the code can be glitched, we now take a closer look at the fuse check.
The code shown in Listing~\ref{listing:condition_check} resembles the fuse check protecting the \texttt{Nv\allowbreak Boot\allowbreak Uart\allowbreak Download()} function.
The two functions \verb|is_fam| (\textit{FailureAnalysisMode}) and \verb|is_ppm| (\textit{PreproductionMode}) always return \verb|0| due to the values in the corresponding fuses.
From line 3, the code normally jumps to line 8, \verb|is_ppm| returns \verb|0|, and the instruction in line 11 does not branch.
Therefore, the routine exits and the \texttt{NvBoot\allowbreak Uart\allowbreak Download()} function is never reached.

However, if a glitch changes the branch direction in line 11 and the branch to \verb|is_fam_or_ppm| is taken, the \texttt{NvBoot\allowbreak Uart\allowbreak Download()} function is subsequently called, since \verb|is_ppm| will still return \verb|0| and the check in line 6 will not branch to \verb|exit|.
Similarly, if the branch in line 3 is skipped, we end up in line 5.
Again, line 6 will not branch since \verb|is_ppm| returns \verb|0|, and we also end up in \texttt{NvBoot\allowbreak Uart\allowbreak Download()}.
As a result of these observations, we hypothesize that the \texttt{NvBoot\allowbreak Uart\allowbreak Download()} method can be activated by injecting a glitch at the correct time during boot.

\subsection{Glitching the BootRom}
\subsubsection{Proof of Concept} \label{par:poc}
\begin{figure}
	\centering
	\includegraphics[width=\linewidth]{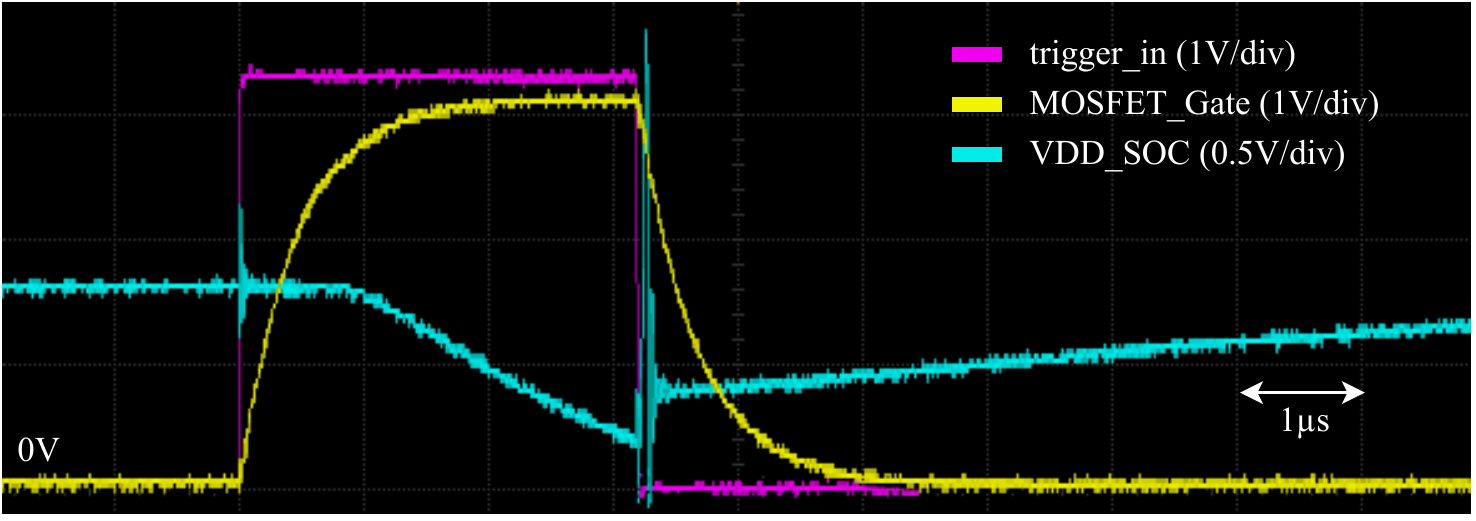}
	\caption{Example glitch of $\approx$\,\SI{3}{\us} length.
	\texttt{trigger\_in} is the \ac{FPGA}'s \ac{GPIO} signal controlling the multiplexer, \texttt{MOSFET\_Gate} is measured at the \ac{MOSFET}'s gate, and \texttt{VDD\_SOC} at the \ac{MOSFET}'s drain.
	\label{fig:glitch_scope}}
\end{figure}

To test the hypothesis of entering the \ac{UART} bootloader by \ac{FI}, we built an \ac{MB2} binary containing assembly code that structurally resembles the code shown in Listing~\ref{listing:condition_check}.
Instead of \verb|is_ppm| and \verb|is_fam|, the binary uses stubs that always return 0.
The binary also includes a stable timing anchor in the form of a \ac{GPIO} pin that is set to high before the critical code section begins.
\texttt{NvBoot\allowbreak Uart\allowbreak Download()} is replaced with a message sent over \ac{UART}.
After running some minor adjustments to the glitch parameters, we saw the success message being printed.
This means that the code can indeed be glitched on the \ac{BPMP}.
Fig.~\ref{fig:glitch_scope} shows the oscilloscope trace of a single glitch.

Knowing that our hypothesis is correct, we will now describe how we attacked the actual boot process.

\subsubsection{Trigger Signal}
The trigger used for the proof of concept was a self-generated signal.
However, for a reliable glitch on the actual target, we need a similar signal during booting, as described in Step~\ref{step_3} of the attack procedure.
In order to find such a signal, we probed signals at different components on the \ac{PCB}, like test points and voltage regulators, using an oscilloscope.
However, none of the tested measuring points revealed trigger signals that potentially correlate with code executed in the early boot process. 
This is confirmed by further analyzing the \ac{BR} code, as no external signals are generated before the relevant code section starts.

Therefore, the only reliable signal at this point is the reset release signal.
In many cases, the reset can not be used as a reliable trigger because there is jitter between reset release and the targeted code section, see Section~\ref{sec:attack_approach:procedure}.
However, since we know from the analyzed \ac{BR} code that there is only minimal code being executed between the reset exception and the fuse check, we can assume that the reset release works as a trigger.
Therefore, we connect both the \verb|trigger_in| and \verb|reset_out| ports of the \ac{FPGA} to a pin on the carrier board that exposes the module's reset signal \verb|RESET_IN|, see Fig.~\ref{fig:setup_overview}.
This way, we can pull \verb|RESET_IN| low to initiate a full system reset and then trigger when we release the signal.

\subsubsection{Search Parameters}
After finding a usable trigger signal, we also need a sensible search space for the glitching parameters as described in Step~\ref{step_4} of the attack procedure.
The parameters to optimize are glitch length and the offset from the trigger.
On the one hand, we want to test as many glitch length/offset combinations as possible, but, on the other hand, including more parameters increases the time to search through all of them.
In the following, we describe our approaches to minimize the search space for both parameters.

\mysubsubsubsection{Glitch Length}
To optimize the glitch length as far as possible, we used our proof of concept \ac{MB2} binary.
By having complete control over the trigger signal, we can run experiments in a very small offset range to find the glitch length interval that leads to the highest success rate.
We marked the beginning and end of the critical code section by setting a \ac{GPIO} pin high in the beginning and low after the section finished.
Then we tried all relevant pulse lengths for every offset in that interval.
First, we determined the range of plausible pulse lengths by setting the offset to 0 and then manually increasing the length of the pulse until the target reliably stops execution.
This is the case for pulse lengths above \SI{13}{\us}.
Subsequently, we tried every possible combination of both parameter ranges and recorded the parameter combination of successful glitches.
After achieving ten successful glitches, we reduced the search intervals by constraining them to the minimum and maximum of the found parameters, plus some tolerance in either direction.
We let the experiment run for eight hours and selected the three pulse lengths that worked most frequently, ending up with the following glitch pulse lengths: \SI{11,30}{\us}, \SI{11,32}{\us}, and \SI{11,34}{\us}.

\mysubsubsubsection{Offset Range} \label{par:offset_range}
For limiting the interval to inject the glitch, we searched for signals and side-channels on the module that would allow us to identify the beginning of subsequent boot stages and operations.
By our capability to execute code during \ac{MB2}, we learned at which point the execution of \ac{MB2} starts.
However, the time span between reset release and the start of \ac{MB2} is \SI{172}{\ms}.
When searching the offset space with the highest possible resolution, one offset step is \SI{20}{\ns} long. 
Therefore there would be $\SI{172}{\ms} / \SI{20}{\ns} = 8600000$ possible offsets.
Assuming that the result can be seen immediately after the glitch, one glitch attempt takes $\SI{172}{\ms} / 2 = \SI{86}{\ms}$ on average.
This means that a full pass would take $(8600000 * \SI{86}{\ms}) / (1000 * 60 * 60 * 24) \approx \SI{8.5}{days}$.
Taking into account that there are three different pulse lengths that should be tested and, more importantly, that the success rate may be below \SI{1}{\percent}, it seems problematic to use the start of \ac{MB2} as the limit for the offset.
Therefore we needed to find an alternative way to minimize the offset interval that needs to be searched.

One crucial step for attacking the X1 is the ability to observe communication between the \ac{eMMC} and the X1 through a debug port~\cite{galauner_glitching_2018}. 
The data observed on this port can be used as a reliable trigger signal.
On the TX2 module, a comparable port is not available.
Therefore, one avenue of research was to look for ways to load the \ac{MB1} image from another device than the \ac{eMMC}, potentially allowing us to obtain a trigger signal during \ac{BR} execution.
We found a thread in the official Nvidia forum that discusses precisely this topic~\cite{nvidia_qspi_boot}.
The original author describes how they use the strapping pins \verb|BOOT_SELECT[2:0]| on their TX2i\footnote{TX2 Industrial: Another Jetson module housing an X2 \ac{SOC} with a different \ac{PCB} and slightly different components} to enter a mode where the \ac{BR} starts looking for a \ac{QSPI} device.

From our analysis of the \ac{BR} in Section~\ref{subsubsec:analyzing_irom}, we know that the strapping pins are not checked before the fuse check.
Furthermore, we know from the leaked source code that the function \texttt{Setup\allowbreak Boot\allowbreak Device()} in file \texttt{nvboot\_\allowbreak coldboot.c} is responsible for checking the strapping pins. 
Since this function is located in the protected section of the \ac{BR}, it will always be executed after the fuse check.
Thus, we can further narrow down the search interval if we can trigger and identify the \ac{QSPI} signals.

The \ac{OEM} Product Design Guide~\cite{x2_oem_guide} explains how the strapping pins are connected on the TX2 module.
The pins are not exposed via the module connector but instead routed into a multiplexer located on the module, with pull-down resistors connected to the lines.
We identified multiple multiplexer \acp{IC} on the module by their chip markings.
By knowing to which pins of the module connector the multiplexer's output is routed and that there are three pull-down resistors nearby, we were able to identify the \ac{IC} with the label \verb|U520| as the most likely candidate.
By applying \SI{3.3}{\V} to the multiplexer inputs at reset release and measuring the signals \verb|QSPI_SCK|, \verb|QSPI_CS_N|, and \verb|QSPI_IO0| (module connector names: G8, H8, and H7, respectively), we were able to observe the \ac{QSPI} clock signal.
This clock signal is only activated if the strapping pin is driven, starting \SI{4.42}{\ms} after reset release.

\subsubsection{Brute Force Search} 
Since we know that the \ac{QSPI} clock will only be activated after the fuse check, this gives us a much tighter upper bound to the offset interval.
Applying the same calculation as earlier in Section~\ref{par:offset_range}, we can approximate that a single run over the search space takes around \SI{8.14}{\minute}.
Since this seemed to be a reasonable time frame, we run the Python control code with the following parameters:
\begin{itemize}
    \item \textbf{Pulse Lengths:} \SI{11.30}{\us}, \SI{11.32}{\us}, \SI{11.34}{\us}
    \item \textbf{Offset Interval:} \SI{0}{\ms} - \SI{4.42}{\ms} with \SI{20}{\ns} steps
\end{itemize}
\noindent
The success criterion was reading the text prompt mentioned in the \ac{BR} analysis via \ac{UART}.
This happened twice, roughly eight hours after starting the experiment. 
The first two successful glitches were at offsets \SI{2.633800}{\ms} and \SI{2.625900}{\ms} with pulse lengths of \SI{11.32}{\us} and \SI{11.34}{\us}, respectively.
To reproduce the result, we constrained the offset to an interval of \SI{50}{\us} around the earlier offset.
In the resulting search interval between \SI{2.585200}{\ms} and \SI{2.634800}{\ms} and the same pulse lengths, the interface can be activated reliably within less than ten seconds.
In most cases, it takes less than two seconds.

\subsubsection{Payload}
To leverage the debug interface's activation to gain code execution on the \ac{BPMP}, the code that should be executed needs to be sent to the X2 over \ac{UART}.
This corresponds to Step~\ref{step_5} of the attack procedure.
While there are no further security measures to authenticate this code, the payload has to adhere to the structure that the \ac{BR} expects.

From the unprotected \ac{BR} binary (see Section~\ref{subsubsec:analyzing_irom}), we can learn how to compose the payload.
Cross-referencing the binary with the leaked code from GitHub indicates that the payload must start with a header called \texttt{NvBoot\allowbreak Uart\_\allowbreak Header}.
It includes the address of the first instruction, the length of the program following the header, and four unique identifications fields that are not relevant for us, as they are not checked.
The actual program code follows the header. 
The last 4 bytes of the binary are a checksum composed by the one's complement of the sum of all preceding bytes.

Following this structure, we built an assembly program that adheres to the structure and sends the \ac{BR}'s protected region via \ac{UART} to the control PC.
That binary could then be saved and imported into Ghidra for further analysis.

\subsection{Decrypting MB1}
After dumping the full \ac{BR}, we started analyzing the code that is responsible for loading and authenticating \ac{MB1}. 
All cryptographic operations are executed by a hardware crypto engine.
This engine also manages the keys used for these operations in several key slots. 
After analyzing the code that configures the hardware crypto engine, we could see that two keys called \verb|FEK| are loaded into temporary key slots.
In analogy to the naming in~\cite{tx2_l4t_spe_armiscreg_header}, we infer the name of these keys as \acp{FEK}.
They are located in an aperture called \verb|MISC| in the SoC. The first key loaded from these registers is named \texttt{FUSE\allowbreak ROM\allowbreak ENCRYPTION\allowbreak NVKEY}.
The second key is, depending on a fuse bit, loaded either from the same source or from registers called \texttt{FUSE\allowbreak ROM\allowbreak ENCRYPTION\allowbreak TESTKEY}~\cite{tx2_l4t_spe_armiscreg_header}.
After the keys are loaded, a bit in an undocumented control register is set to protect these keys from being read again.

Next, several encrypted keys are read from the topmost \SI{4}{KiB} of the \ac{BR} memory region into other hardware key slots.
These keys are then decrypted using \ac{AES} in \ac{CBC} mode with the first \ac{FEK} and an \ac{IV} of all zeros.
The decrypted keys are internal to Nvidia and are used to decrypt assets loaded later in the boot chain.
After that, the \ac{BR} can derive keys used to implement a secure boot mechanism for \acp{ODM} using the second \ac{FEK}, if this feature has been enabled by an \ac{ODM} using a fuse bit.

The \ac{BR} then proceeds to load the \ac{MB1} from non-volatile storage and to decrypt and authenticate it using the derived Nvidia \ac{MB1} keys.
If an \ac{ODM} wants to enable authentication/encryption for their software, i.e., for \ac{MB2} and later stages, \ac{MB1} is also encrypted and signed using the \ac{ODM}'s key material to protect the configuration of \ac{MB1} through the \ac{BCT}.
In such cases, \ac{MB1} is first authenticated/decrypted using the \ac{ODM} key material, then using the Nvidia key material.
Before jumping to the fully authenticated and decrypted \ac{MB1}, the corresponding key slot containing the used keys for decryption is cleared to prevent it from being used by code loaded in later boot stages. 
All other key slots are kept intact and are used by \ac{MB1} to decrypt other boot loaders, metadata, and microcode.

Once we recognized that we were missing the \acp{FEK} to decrypt \ac{MB1}, we enhanced the payload used to dump the \ac{BR} to extract the registers containing the \acp{FEK} as well.
The code path responsible for performing the \ac{UART} pre-production serial boot that we entered using \ac{FI} does not disable the readout of the \acp{FEK}.
Therefore, we were able to dump the keys and use them together with the previously dumped \ac{BR} to derive the \ac{MB1} decryption key, and decrypt the \ac{MB1} binary.

\section{Discussion\label{sec:discussion}}

\subsection{Attack Applicability and Impact}
Our work has shown that the boot processor on the Nvidia Tegra X2 \ac{SOC} is susceptible to voltage \ac{FI}.
We uncovered a hidden \ac{UART} bootloader and forced its activation by \ac{FI}.
This bootloader allows us to execute code in the Secure/TZ mode.
Consequently, we could extract the full \ac{BR} code, as well as all fuses and registers holding cryptographic keys.
Using the recovered key material, we were able to decrypt the second-stage bootloader (\ac{MB1}).
Together with the extracted fuse values, an attacker would now be able to decrypt \ac{OEM}'s binaries of later boot stages.
These capabilities form the perfect steppingstones to develop further attacks against a particular product. 

Next to these insights, the \ac{UART} bootloader would enable an adversary to take control over the boot process and the entire system.
They could load their own unauthenticated code or disable security settings and then mimic the original system's behavior.
Due to the short time of only a few seconds to a successful exploit, the \ac{FI} can be conducted at each boot of the system without a significant increase in boot time.
Moreover, in contrast to other more complicated \ac{FI} techniques using lasers or \ac{EM} radiation, this voltage \ac{FI} capability can be persisted using a hardware implant.
This has been demonstrated to be feasible, e.g., by Team Xecuter with their Nintendo Switch implant~\cite{team_xecuter_implant}.

Our attack has shown that simple things like a bootloader that should only be used by Nvidia and is not officially documented can be leveraged to break the system's entire security.
Devices and products of other manufacturers that are not hardened against voltage \ac{FI} will likely have similar weaknesses, which can be exploited easily with readily available equipment.

\subsection{Potential Mitigations\label{sec:discussion:mitigations}}
To protect \acp{SOC} against voltage \ac{FI} attacks, we distinguish between mitigations that can be applied to prevent attacks on existing architectures and mitigations directly implemented into future chip generations.

\subsubsection{Protecting Existing SoCs\label{sec:discussion:mitigations:existing}}
After disclosing our findings to Nvidia, they proposed measures that can harden a device against voltage \ac{FI} attacks.
Since attackers need physical access for \ac{FI} and similar attacks, \ac{OEM} designers can increase the attack effort by adding board-level mitigations.
For instance, heat- and solvent-resistant epoxy and non-removable metal shielding can prevent access to sensitive voltage rails and the removal of decoupling capacitors.
Furthermore, embedding capacitive metal layers or decoupling capacitors into the \ac{PCB} could reduce the effectiveness or accuracy of \ac{FI} to a certain extent.
Nevertheless, if the adversary only wants to exploit a single device, they might also consider developing a custom \ac{PCB} for the \ac{SOC} that facilitates access to the signal and omits decoupling capacitors.

\subsubsection{Protecting New SoC Versions\label{sec:discussion:mitigations:new}}
The threat of fault injection for gaining adversarial code execution can be tackled from different directions.
One could try to detect malicious voltage drops/glitches, and as a consequence, shut down the system to prevent further damage.
Alternatively, one could try to prevent faulty execution in the presence of glitches, for instance, by introducing redundancy.
These approaches imply changes in hardware or software design.
Since neither approach can eradicate the chance of successful glitch injection completely, both approaches should be employed at the same time.

\mysubsubsubsection{Hardware-based Mitigations}
Voltage monitoring circuits -- as commonly implemented in modern smartcards -- could help to detect glitches.
Nvidia has recently patented a cross-domain voltage glitch detection circuit, which can be implemented into an \ac{SOC}~\cite{rajpathak_cross_2020}.
The main idea is that an \ac{SOC} contains multiple voltage rails.
Therefore, circuits in independent voltage domains can monitor voltage levels in other domains, and if there is a glitch on a specific rail, assert an alert signal.
Although we think this is a promising approach, it should be kept in mind that there might exist voltage glitch shapes that can cause faulty behavior but can not be detected by a particular protection circuit.

Fully integrating the voltage regulators into the \ac{SOC} could be another solution.
However, faults can not only be induced by glitching the supply voltage.
In the past couple of years, \ac{EM} fault injection techniques against modern CPUs have been examined to inject faults in a targeted and contactless way~\cite{trouchkine_fault_2020, trouchkine_electromagnetic_2021}.
Consequently, a holistic view is necessary to prevent all kinds of fault injection attacks that can manipulate the behavior of the target device.

\mysubsubsubsection{Software-based Mitigations}
Hardening the \ac{BR} might be another option to prevent the adversary from gaining code execution.
However, this is a complex task since the characteristics and potentials of faults are not well understood.
Particularly, there is no model which covers all possible faults.
Nevertheless, Riscure and ARM propose countermeasures that can decrease the probability of successful attacks~\cite{witteman_riscure_2018, ban_arm_2020}.
For instance, constants with large hamming distances can complicate flipping one valid value to another, and double-checks can protect branch conditions.
Moreover, loop integrity checks make sure that the loop exits as intended, and a global counter can be used to monitor the program flow and detect anomalies.
For assessing software countermeasures against fault attacks, different simulation-based frameworks have been proposed~\cite{holler_qemubased_2015, schirmeier_fail_2015}.

It should be noted that Nvidia is seemingly aware of these countermeasures. 
We saw most of these countermeasures being used the leaked X1+ "Mariko" \ac{BR} source code.
Unfortunately, they are not employed in the X2.
The general approach of software-based mitigations might be promising as they can also protect against fault attacks other than voltage glitching.
\section{Conclusion}

Starting with the knowledge that the Nvidia Tegra X1 \ac{SOC} is susceptible to voltage \ac{FI}, we have investigated the boot security on its successor, the Tegra X2.
We have shown that a hidden bootloader, which allows code execution with the highest privileges, can be enabled using voltage \ac{FI}.
Using this capability, we were able to extract the entire bootloader's code and decryption keys for later boot stages.
Furthermore, we described that an adversary could use the bootloader for injecting unauthentic code to bypass trusted code execution on the system.
After disclosing our findings to Nvidia, they confirmed the vulnerability and stated that future device generations contain mitigations against \ac{FI} attacks.
As devices of other manufacturers potentially are vulnerable in a similar manner, we have discussed countermeasures to mitigate our attack.
Manufacturers and designers should not forget about seemingly simple hardware attacks that have been around for already more than two decades.

\ifthenelse{\boolean{cameraready}}{
\section*{Acknowledgment}
The work described in this paper has been supported by the Einstein Foundation in form of an Einstein professorship (EP-2018-480).
}{}

\bibliographystyle{./IEEEtran}
\bibliography{./bib/IEEEabrv,./bib/references}

\begin{thebibliography}{10}
\providecommand{\url}[1]{#1}
\csname url@samestyle\endcsname
\providecommand{\newblock}{\relax}
\providecommand{\bibinfo}[2]{#2}
\providecommand{\BIBentrySTDinterwordspacing}{\spaceskip=0pt\relax}
\providecommand{\BIBentryALTinterwordstretchfactor}{4}
\providecommand{\BIBentryALTinterwordspacing}{\spaceskip=\fontdimen2\font plus
\BIBentryALTinterwordstretchfactor\fontdimen3\font minus
  \fontdimen4\font\relax}
\providecommand{\BIBforeignlanguage}[2]{{%
\expandafter\ifx\csname l@#1\endcsname\relax
\typeout{** WARNING: IEEEtran.bst: No hyphenation pattern has been}%
\typeout{** loaded for the language `#1'. Using the pattern for}%
\typeout{** the default language instead.}%
\else
\language=\csname l@#1\endcsname
\fi
#2}}
\providecommand{\BIBdecl}{\relax}
\BIBdecl

\bibitem{bar-el_sorcerer_2006}
H.~{Bar-El}, H.~Choukri, D.~Naccache, M.~Tunstall, and C.~Whelan, ``The
  {{Sorcerer}}'s {{Apprentice Guide}} to {{Fault Attacks}},'' \emph{Proceedings
  of the IEEE}, vol.~94, no.~2, pp. 370--382, Feb. 2006.

\bibitem{koemmerlin_design_1999}
O.~Koemmerling and M.~G. Kuhn, ``Design {{Principles}} for
  {{Tamper}}-{{Resistant Smartcard Processors}},'' \emph{Smartcard 99}, pp.
  9--20, 1999.

\bibitem{boneh_importance_2001}
D.~Boneh, R.~A. DeMillo, and R.~J. Lipton, ``On the {{Importance}} of
  {{Eliminating Errors}} in {{Cryptographic Computations}},'' \emph{Journal of
  Cryptology}, vol.~14, no.~2, pp. 101--119, Mar. 2001.

\bibitem{timmers_controlling_2016}
N.~Timmers, A.~Spruyt, and M.~Witteman, ``Controlling {{PC}} on {{ARM Using
  Fault Injection}},'' in \emph{2016 {{Workshop}} on {{Fault Diagnosis}} and
  {{Tolerance}} in {{Cryptography}} ({{FDTC}})}, Aug. 2016, pp. 25--35.

\bibitem{lu_injecting_2019}
\BIBentryALTinterwordspacing
Y.~Lu. (2019) Injecting {{Software Vulnerabilities}} with {{Voltage
  Glitching}}. [Online]. Available: \url{https://arxiv.org/abs/1903.08102}
\BIBentrySTDinterwordspacing

\bibitem{vasselle_laserinduced_2020}
A.~Vasselle, H.~Thiebeauld, Q.~Maouhoub, A.~Morisset, and S.~Ermeneux,
  ``Laser-{{Induced Fault Injection}} on {{Smartphone Bypassing}} the {{Secure
  Boot}}-{{Extended Version}},'' \emph{IEEE Transactions on Computers},
  vol.~69, no.~10, pp. 1449--1459, Oct. 2020.

\bibitem{abdellatif_silicontoaster_2020}
K.~M. Abdellatif and O.~H{\'e}riveaux, ``{{SiliconToaster}}: {{A Cheap}} and
  {{Programmable EM Injector}} for {{Extracting Secrets}},'' in \emph{2020
  {{Workshop}} on {{Fault Detection}} and {{Tolerance}} in {{Cryptography}}
  ({{FDTC}})}, Sep. 2020, pp. 35--40.

\bibitem{trouchkine_electromagnetic_2021}
T.~Trouchkine, S.~K. Bukasa, M.~Escouteloup, R.~Lashermes, and G.~Bouffard,
  ``{Electromagnetic} {Fault} {Injection} against a {Complex} {{CPU}}, toward
  {New} {Micro-Architectural} {Fault} {Models},'' \emph{Journal of
  Cryptographic Engineering}, Mar. 2021.

\bibitem{djellid-ou_supply_2006}
A.~{Djellid-Ouar}, G.~Cathebras, and F.~Bancel, ``Supply voltage glitches
  effects on {{CMOS}} circuits,'' in \emph{International {{Conference}} on
  {{Design}} and {{Test}} of {{Integrated Systems}} in {{Nanoscale
  Technology}}, 2006. {{DTIS}} 2006.}\hskip 1em plus 0.5em minus 0.4em\relax
  {IEEE}, 2006, pp. 257--261.

\bibitem{galauner_glitching_2018}
\BIBentryALTinterwordspacing
A.~Galauner. (2018) \BIBforeignlanguage{en}{Glitching the {{Switch}}}. Video.
  [Online]. Available:
  \url{https://media.ccc.de/v/c4.openchaos.2018.06.glitching-the-switch}
\BIBentrySTDinterwordspacing

\bibitem{nvidiacor_autonomous_2021}
\BIBentryALTinterwordspacing
{Nvidia Corp.} (2021) {NVIDIA DRIVE PX - Scalable AI Supercomputer For
  Autonomous Driving}. [Online]. Available:
  \url{https://www.nvidia.com/content/nvidiaGDC/sg/en_SG/self-driving-cars/drive-px/}
\BIBentrySTDinterwordspacing

\bibitem{nvidiacor_tesla_2016}
\BIBentryALTinterwordspacing
------. (2016) Tesla {{Motors}}' {{Self}}-{{Driving Car}} ``{{Supercomputer}}''
  {{Powered}} by {{NVIDIA DRIVE PX}} 2 {{Technology}}. [Online]. Available:
  \url{https://blogs.nvidia.com/blog/2016/10/20/tesla-motors-self-driving/}
\BIBentrySTDinterwordspacing

\bibitem{daimlerag_glance_2018}
\BIBentryALTinterwordspacing
{Daimler AG}. (2018) {At} a {Glance}: {{The}} {Key} {Data} on {{MBUX}}.
  [Online]. Available:
  \url{https://media.daimler.com/marsMediaSite/ko/en/32705799}
\BIBentrySTDinterwordspacing

\bibitem{abuelsamid_hyundai_2020}
\BIBentryALTinterwordspacing
S.~Abuelsamid. (2020) Hyundai {{Adopts Nvidia Parker}} `{{System On A Chip}}'
  {{For Infotainment}}. [Online]. Available:
  \url{https://www.forbes.com/sites/samabuelsamid/2020/11/11/hyundai-motor-group-adopts-nvidia-parker-soc-for-infotainment/}
\BIBentrySTDinterwordspacing

\bibitem{oflynn_chipwhisperer_2014}
C.~O'Flynn and Z.~D. Chen, ``{{ChipWhisperer}}: {{An Open}}-{{Source Platform}}
  for {{Hardware Embedded Security Research}},'' in \emph{Constructive
  {{Side}}-{{Channel Analysis}} and {{Secure Design}}}.\hskip 1em plus 0.5em
  minus 0.4em\relax {Springer International Publishing}, 2014, pp. 243--260.

\bibitem{bozzato_shaping_2019}
C.~Bozzato, R.~Focardi, and F.~Palmarini, ``Shaping the {{Glitch}}:
  {{Optimizing Voltage Fault Injection Attacks}},'' \emph{IACR Transactions on
  Cryptographic Hardware and Embedded Systems}, pp. 199--224, Feb. 2019.

\bibitem{chen_voltpillager_2020}
Z.~Chen, G.~Vasilakis, K.~Murdock, E.~Dean, D.~Oswald, and F.~D. Garcia,
  ``{{VoltPillager}}: {{Hardware}}-based {Fault} {Injection} {Attacks} against
  {{Intel SGX Enclaves}} using the {{SVID}} {Voltage} {Scaling} {Interface},''
  in \emph{30th {USENIX} Security Symposium ({USENIX} Security 21)}.\hskip 1em
  plus 0.5em minus 0.4em\relax {USENIX} Association, Aug. 2021.

\bibitem{murdock_plundervolt_2020}
K.~Murdock, D.~Oswald, F.~D. Garcia, J.~V. Bulck, F.~Piessens, and D.~Gruss,
  ``Plundervolt: {{How}} a {{Little Bit}} of {{Undervolting Can Create}} a
  {{Lot}} of {{Trouble}},'' \emph{IEEE Security \& Privacy}, vol.~18, no.~5,
  pp. 28--37, Sep. 2020.

\bibitem{qiu_voltjockey_2020}
P.~Qiu, D.~Wang, Y.~Lyu, R.~Tian, C.~Wang, and G.~Qu, ``{{VoltJockey}}: {{A New
  Dynamic Voltage Scaling}} based {{Fault Injection Attack}} on {{Intel
  SGX}},'' \emph{IEEE Transactions on Computer-Aided Design of Integrated
  Circuits and Systems}, pp. 1--1, 2020.

\bibitem{oflynn_fault_2016}
\BIBentryALTinterwordspacing
C.~O'Flynn, ``Fault {{Injection}} using {{Crowbars}} on {{Embedded Systems}},''
  \emph{IACR Cryptol. ePrint Arch.}, 2016. [Online]. Available:
  \url{https://eprint.iacr.org/2016/810}
\BIBentrySTDinterwordspacing

\bibitem{milburn_ecu_glitches_2018}
A.~Milburn, N.~Timmers, N.~Wiersma, R.~Pareja, and S.~Cordoba,
  ``\BIBforeignlanguage{en}{There {Will} {Be} {Glitches}: {Extracting} and
  {Analyzing} {Automotive} {Firmware} {Efﬁciently}},''
  \emph{\BIBforeignlanguage{en}{{Black Hat USA}}}, 2018.

\bibitem{herrewegen_fill_2021}
J.~V. den Herrewegen, D.~Oswald, F.~D. Garcia, and Q.~Temeiza, ``Fill your
  {{Boots}}: {{Enhanced Embedded Bootloader Exploits}} via {{Fault Injection}}
  and {{Binary Analysis}},'' \emph{IACR Transactions on Cryptographic Hardware
  and Embedded Systems}, pp. 56--81, 2021.

\bibitem{ps3_hack}
\BIBentryALTinterwordspacing
{Nate Lawson}. How the {PS3} {Hypervisor} was {Hacked}. [Online]. Available:
  \url{https://rdist.root.org/2010/01/27/how-the-ps3-hypervisor-was-hacked/}
\BIBentrySTDinterwordspacing

\bibitem{xbox_hack}
\BIBentryALTinterwordspacing
{Razkar}. The {Reset} {Glitch} {Hack} - {A} {New} {Exploit} on {Xbox} 360.
  [Online]. Available:
  \url{http://www.logic-sunrise.com/news-341321-the-reset-glitch-hack-a-new-exploit-on-xbox-360-en.html}
\BIBentrySTDinterwordspacing

\bibitem{switch_tx1_dieshots}
\BIBentryALTinterwordspacing
{TechInsights Inc.} {Nintendo} {Switch} {Teardown}. [Online]. Available:
  \url{https://www.techinsights.com/blog/nintendo-switch-teardown}
\BIBentrySTDinterwordspacing

\bibitem{szekely_qyriad_2021}
\BIBentryALTinterwordspacing
M.~Szekely. (2021, Jul.) Qyriad/fusee-launcher. [Online]. Available:
  \url{https://github.com/Qyriad/fusee-launcher/blob/master/report/fusee_gelee.md}
\BIBentrySTDinterwordspacing

\bibitem{rousseltarbouriech2019methodically}
\BIBentryALTinterwordspacing
G.~T. H. G.~I. Roussel-Tarbouriech, N.~Menard, T.~True, T.~Vi, and Reisyukaku.
  (2019) {Methodically} {Defeating} {Nintendo} {Switch} {Security}. [Online].
  Available: \url{https://arxiv.org/abs/1905.07643}
\BIBentrySTDinterwordspacing

\bibitem{console_security_switch}
\BIBentryALTinterwordspacing
plutoo, derrek, and naehrwert. (2017) \BIBforeignlanguage{en}{Console security
  {{Switch}}}. Video. [Online]. Available:
  \url{https://media.ccc.de/v/34c3-8941-console_security_-_switch}
\BIBentrySTDinterwordspacing

\bibitem{nvidiacor_jetson_2017}
\BIBentryALTinterwordspacing
{Nvidia Corp.} (2017) Jetson {{TX2 Module}}. [Online]. Available:
  \url{https://developer.nvidia.com/embedded/jetson-tx2}
\BIBentrySTDinterwordspacing

\bibitem{nvidiacor_harness_2017}
\BIBentryALTinterwordspacing
------. (2017) Harness {{AI}} at the {{Edge}} with the {{Jetson TX2 Developer
  Kit}}. [Online]. Available:
  \url{https://developer.nvidia.com/embedded/jetson-tx2-developer-kit}
\BIBentrySTDinterwordspacing

\bibitem{tx2_bootflow}
\BIBentryALTinterwordspacing
------. {Jetson TX2 Boot Flow}. [Online]. Available:
  \url{https://docs.nvidia.com/jetson/l4t/index.html#page/Tegra%20Linux%20Driver%20Package%20Development%20Guide/bootflow_tx2.html}
\BIBentrySTDinterwordspacing

\bibitem{x2_trm}
------, \emph{Technical Reference Manual Nvidia Parker Series SoC, v1.0p}, Jun.
  2017.

\bibitem{chipfail_repo}
\BIBentryALTinterwordspacing
{Thomas Roth}. {Chipfail-glitcher}. GitHub. [Online]. Available:
  \url{https://github.com/chipfail/chipfail-glitcher}
\BIBentrySTDinterwordspacing

\bibitem{gayou_sgayou_2020}
\BIBentryALTinterwordspacing
S.~Gayou. (2020) Sgayou/rbasefind. GitHub. [Online]. Available:
  \url{https://github.com/sgayou/rbasefind}
\BIBentrySTDinterwordspacing

\bibitem{nvidiacor_openvreg_2011}
\BIBentryALTinterwordspacing
{Nvidia Corp.}, \emph{{{OpenVReg}}: {{Open Voltage Regulator Specifcation}}},
  2011. [Online]. Available:
  \url{https://international.download.nvidia.com/openvreg/openvreg-type2-plus-1-specification.pdf}
\BIBentrySTDinterwordspacing

\bibitem{nvidia_l4t}
\BIBentryALTinterwordspacing
------, \emph{{L}inux {F}or {T}egra}, 32nd~ed., Nvidia Corporation, Feb. 2021.
  [Online]. Available: \url{https://developer.nvidia.com/embedded/linux-tegra}
\BIBentrySTDinterwordspacing

\bibitem{x2_driver_guide}
\BIBentryALTinterwordspacing
------, \emph{{Tegra Linux Driver Package Development Guide 27.1}}, Mar. 2017.
  [Online]. Available:
  \url{https://docs.nvidia.com/jetson/archives/l4t-archived/l4t-271/index.html}
\BIBentrySTDinterwordspacing

\bibitem{tx2_l4t_spe_armiscreg_header}
\BIBentryALTinterwordspacing
------. {Jetson} {Sensor} {Processing} {Engine} ({SPE}). Source code. [Online].
  Available:
  \url{https://developer.nvidia.com/embedded/L4T/r32_Release_v5.0/sources/T186/l4t_rt_aux_cpu_src.tbz2}
\BIBentrySTDinterwordspacing

\bibitem{zerospacenx_2021}
\BIBentryALTinterwordspacing
highmirai. (2021) Zerospace-nx/switch-bootroms. GitHub. [Online]. Available:
  \url{https://github.com/zerospace-nx/switch-bootroms}
\BIBentrySTDinterwordspacing

\bibitem{tegra_2021}
\BIBentryALTinterwordspacing
{Wikipedia}. (2021) Tegra. [Online]. Available:
  \url{https://en.wikipedia.org/wiki/Tegra#Tegra_X1}
\BIBentrySTDinterwordspacing

\bibitem{nvidia_fskp_patent}
\BIBentryALTinterwordspacing
J.~Huang, P.~Chou, and A.~WOO, ``Secure provisioning of semiconductor chips in
  untrusted manufacturing factories,'' US Patent US10\,387\,653B2, 2019.
  [Online]. Available: \url{https://patents.google.com/patent/US10387653B2/en}
\BIBentrySTDinterwordspacing

\bibitem{abdel_tawab_fskp_keys}
\BIBentryALTinterwordspacing
R.~Abdel-Tawab. (2019) {Android{\_}bootable{\_}bootloader{\_}cboot-t186}.
  GitHub. [Online]. Available:
  \url{https://github.com/Rashed97/android_bootable_bootloader_cboot-t186/blob/lineage-16.0/hwinc-t18x/nvboot_se_aes.h#L145-L221}
\BIBentrySTDinterwordspacing

\bibitem{nvidia_qspi_boot}
\BIBentryALTinterwordspacing
{Nvidia Corp.} (2019) {Boot source other than eMMC or USB}. Forum entry.
  [Online]. Available:
  \url{https://forums.developer.nvidia.com/t/boot-source-other-than-emmc-or-usb/81621}
\BIBentrySTDinterwordspacing

\bibitem{x2_oem_guide}
------, \emph{{OEM Product Design Guide Nvidia} {J}etson {TX}2 {S}eries}, May
  2020.

\bibitem{team_xecuter_implant}
\BIBentryALTinterwordspacing
T.~Xecuter. (2021) {SX} {Core} {Manual}. [Online]. Available:
  \url{https://sx.xecuter.com/download/manuals/sxcore/[EN]_SX_Core_v1.1.pdf}
\BIBentrySTDinterwordspacing

\bibitem{rajpathak_cross_2020}
\BIBentryALTinterwordspacing
K.~Rajpathak and T.~Raja, ``Cross domain voltage glitch detection circuit for
  enhancing chip security,'' US Patent US20\,200\,285\,780A1, Sep., 2020.
  [Online]. Available:
  \url{https://patents.google.com/patent/US20200285780A1/en}
\BIBentrySTDinterwordspacing

\bibitem{trouchkine_fault_2020}
T.~Trouchkine, G.~Bouffard, and J.~Cl{\'e}di{\`e}re, ``Fault {{Injection
  Characterization}} on {{Modern CPUs}},'' in \emph{Information {{Security
  Theory}} and {{Practice}}}, ser. Lecture {{Notes}} in {{Computer
  Science}}.\hskip 1em plus 0.5em minus 0.4em\relax {Cham}: {Springer
  International Publishing}, 2020, pp. 123--138.

\bibitem{witteman_riscure_2018}
\BIBentryALTinterwordspacing
M.~Witteman. (2018) Riscure: {{Secure Application Programming}} in the
  {{Presence}} of {{Side Channel Attacks}}. [Online]. Available:
  \url{https://www.riscure.com/uploads/2018/11/201708_Riscure_Whitepaper_Side_Channel_Patterns.pdf}
\BIBentrySTDinterwordspacing

\bibitem{ban_arm_2020}
\BIBentryALTinterwordspacing
T.~Ban. (2020) Arm {{Ltd}}.: {{Trusted Firmware M}}. Presentation slides.
  [Online]. Available:
  \url{https://www.trustedfirmware.org/docs/TF-M_fault_injection_mitigation.pdf}
\BIBentrySTDinterwordspacing

\bibitem{holler_qemubased_2015}
A.~H{\"o}ller, A.~Krieg, T.~Rauter, J.~Iber, and C.~Kreiner, ``{{QEMU}}-{{Based
  Fault Injection}} for a {{System}}-{{Level Analysis}} of {{Software
  Countermeasures Against Fault Attacks}},'' in \emph{2015 {{Euromicro
  Conference}} on {{Digital System Design}}}, Aug. 2015, pp. 530--533.

\bibitem{schirmeier_fail_2015}
H.~Schirmeier, M.~Hoffmann, C.~Dietrich, M.~Lenz, D.~Lohmann, and O.~Spinczyk,
  ``{{FAIL}}*: {{An Open}} and {{Versatile Fault}}-{{Injection Framework}} for
  the {{Assessment}} of {{Software}}-{{Implemented Hardware Fault
  Tolerance}},'' in \emph{2015 11th {{European Dependable Computing
  Conference}} ({{EDCC}})}.\hskip 1em plus 0.5em minus 0.4em\relax {IEEE}, Sep.
  2015, pp. 245--255.

\end{thebibliography}

\end{document}